\documentclass{aa}  
\usepackage{graphicx}
\usepackage{txfonts}
\usepackage[table]{xcolor}
\usepackage{subfig}
\usepackage{hyperref}
\usepackage[normalem]{ulem}
\newcommand{\rev}[1]{{\textcolor{black}{#1}}}

\newcommand{\revA}[1]{{\textcolor{black}{#1}}}
\definecolor{ochre}{rgb}{0.8, 0.47, 0.13}

\newcommand{\revAA}[1]{{\textcolor{black}{#1}}}

\begin{document} 

   \title{Physical conditions in PDRs revealed by IGRINS ro-vibrational H$_2$ observations }

   \author{ A. Piluso \inst{1}
   \and  V. Maillard \inst{2}
   \and R. Meshaka \inst{3}
   \and E. Bron \inst{1}
   \and F. Le Petit \inst{1}  
   \and K. F. Kaplan \inst{4}
   \and P. Palud \inst{5}
   \and E. Habart \inst{6}
   \and E. Roueff \inst{1}
   \and J. Le Bourlot \inst{1}
   \and A. Doussot \inst{1}
   \and D. Languignon \inst{1}
   \and N. Moreau \inst{1}
          }

   \institute{LUX, Observatoire de Paris, PSL Research University, CNRS, Sorbonne Universités,  92190 Meudon, France
         \and
         \rev{Instituto de Física Fundamental (CSIC), Calle Serrano 121, 28006, Madrid, Spain}
         \and
                       Astronomy Unit, School of Physics and Astronomy, Queen Mary University of London, London E1 4NS, UK
           \and
                        University of Texas at Austin, Department of Astronomy, 2515 Speedway, Stop C1400, Austin, TX 78712-1205, USA
          \and 
             Université Paris Cité, CNRS, Astroparticule et Cosmologie, 75013 Paris, France
             \and
            Université Paris-Saclay, CNRS, Institut d’Astrophysique Spatiale, 91405 Orsay, France
             }

   \date{February 2026}

  \abstract 
   {Enhanced spatial and spectral resolutions in the near-infrared provided by new instruments allow for unprecedented observations in photodissociation regions (PDRs) of numerous ro-vibrational H$_2$ emission lines, which are key tracers of the chemical and physical stratification of PDRs. Recent spectral surveys across multiple PDRs, combined with the latest advancements in PDR modeling, make possible a significant step forward in understanding the physics and chemistry of PDRs. }
   {We compare H$_2$ high ro-vibrational observations of 5 PDRs \rev{(S140, IC63, Horsehead Nebula, NGC 2023 and Orion Bar)} obtained with the IGRINS spectrograph to PDR models produced with the Meudon PDR code, in order to estimate the physical conditions and to constrain the physical and chemical processes that govern PDRs.}
   {We use newly flux-calibrated and extinction-corrected IGRINS H$_2$ observations \rev{covering the 1.45 to 2.45 $\mu m$ range at a resolution of 45000}, and adjust Meudon PDR models using both a simple $\chi^2$ minimization approach and \revA{a robust method designed to thoroughly explore the physical parameter space using Bayesian inversion followed by posterior exploration via an advanced MCMC technique with both global and local exploration}. The PDR models use specific incident \revAA{far ultraviolet (FUV)} spectra for each PDRs based on theoretical stellar spectra from the Pollux database.}
   {The Meudon PDR code is able to reproduce more than 85\% of the observed H$_2$ ro-vibrational line intensities within a factor of two. We find that (1) a realistic modeling of the incident \rev{FUV} field (both in term of geometry and of spectral shape) and (2) the inclusion of recent data on collisional de-excitation rates for high vibrational levels of H$_2$ are key to this success. H$_2$ ro-vibrational emission lines are found to provide good constraints on both the thermal pressure, $P_\mathrm{th}$ and the incident \rev{FUV} field strength, $G_0$, although with a non-negligible remaining degeneracy in most PDRs. Additional constraints, such as the spatial scales of the PDR derived from ALMA or JWST observations, are found to be able to lift this degeneracy. For low excitation PDRs, the observations provide evidence that nascent H$_2$ molecules formed on grains have relatively low rotational energy and high vibrational energy, as predicted by theoretical and experimental studies of the surface formation of H$_2$.
 The values of $P_\mathrm{th}$ and $G_0$ found for this small sample of PDRs are consistent with a $P_\mathrm{th}$-$G_0$ relationship comparable to that previously derived from rotationally-excited CO observations and from atomic tracers, and proposed to relate to photo-evaporation processes.}
{The ro-vibrational lines of H$_2$, \rev{coming from high vibrational levels (up to $v=13$) in five typical PDRs,} are \rev{successfully} reproduced by \rev{stationary} state-of-the-art PDR models and constrain the physical conditions \rev{near the dissociation front}. \rev{For the first time, they also provide} observational constraints on the level population of H$_2$ at formation on grains. Highly excited vibrational H$_2$ lines, detected only by IGRINS, can complement JWST observations in the study of PDRs by being sensitive to the specific processes populating highly excited levels.}
     

   \maketitle
\section{Introduction}

Constraining the physical mechanisms that regulate star formation in galaxies remains a major challenge in astrophysics. Although significant progress has been made in understanding the star formation laws and the dynamical properties of giant molecular clouds, the local processes involved are still poorly understood \citep{Krumholz2014}. In particular, mechanical processes such as stellar winds and supernova explosions, along with radiative processes driven by UV photons emitted by young massive stars, profoundly affect the interstellar gas. This can lead to compression that either triggers the formation of a new generation of stars or, conversely, heats, disperses the gas and destroys its molecular content rendering it unsuitable for star formation.

The layer of gas where far-UV photons (6–13.6 eV) from young massive stars penetrate the neutral medium and govern the transition from molecular to atomic gas is known as a photo-dissociation region (PDR) \citep{Tielens1985}. Located at the surface of molecular clouds, in particular on the surfaces surrounding H\textsc{ii} regions produced by young massive stars within or close to the cloud, these regions play a crucial role in the interstellar medium (ISM). They delineate the boundary between gas that is unable to form stars and the cold molecular gas that can. Heated to several hundred Kelvins by the photoelectric effect on grains \rev{and PAH} and by collisional de-excitation of H$_2$, PDRs emit in numerous atomic and molecular lines, in addition to their dust continuum emission, making them unique laboratories for studying the impact of radiative feedback from young stars on their parent molecular clouds.

Photodissociation regions have been extensively studied since the 1980s. The classical view describes PDRs as strongly stratified regions with successive atomic-to-molecular transitions: starting with the H/H$_2$ transition, followed by the C$^+$/CO transition, ultimately leading to cold molecular gas where complex chemistry occurs both in the gas phase and on grain surfaces \citep{Hollenbach1997, Hollenbach1999, Wolfire2022}. The main diagnostics used to probe the physical conditions in PDRs have been atomic fine-structure lines such as \rev{[C$\textsc{ii}]$} at 158 $\mu$m, $[\mathrm{O}\textsc{i}]$ at 63 and 145 $\mu$m, and $[\mathrm{C}\textsc{i}]$ at 370 and 609 $\mu$m, along with molecular rotational transitions (CO, pure rotational H$_2$, HCO$^+$,...) and a few vibrational transitions of H$_2$ observable from space and the ground.

Some authors proposed that PDRs consist of dense clumps embedded within a more diffuse inter-clump medium. In this scenario, atomic line emission predominantly arises from the low density gas whereas molecular excited emission is produced at the surfaces of dense clumps irradiated by UV photons that penetrate deeply into the clumpy structure \citep{Meixner1993, Hogerheijde1995}. However, this view has been challenged in recent years by Herschel and ALMA \rev{(e.g. \citealt{Joblin2018,Goicoechea2016})}, and confirmed with JWST observations which revealed that PDRs exhibit filamentary morphologies \rev{\citep{Berne2022,Peeters2024,Habart2024}}, corresponding to edge-on portions of a well-defined, thin, but corrugated emitting surface layer, where the gas is compressed to relatively high thermal pressures (P$_{\text{th}}$/k $\simeq$ a few $10^7$ - $10^8$ K cm$^{-3}$). Recent H$_2$ JWST observations show a filamentary morphology in PDRs as for example in the Orion Bar \citep{Habart2024} and the Horsehead PDR \citep{Abergel2024,Zannese2025}. It has been proposed that this compression is caused by a rocket effect induced by the photo-evaporation driven at the ionization front by the intense UV radiation fields of nearby O and B stars \citep{Goicoechea2016, Bron2018}. Furthermore, the latest observations of the Horsehead Nebula in pure rotational $\text{H}_2$ using the JWST MIRI-MRS and NIRSpec instruments have enabled the estimation of the pressure and temperature of the rotational $\text{H}_2$ emitting layer. \citet{Zannese2025} finds a pressure exceeding $6 \times 10^6\ \text{K}\ \text{cm}^{-3}$ which is significantly greater than the pressure in the ionized medium and $T_{\text{gas}} = 500 \pm 20 $ K in the first filament at the edge of the PDR.
A detailed analysis of the molecular lines emitted in PDRs and in particular of the high-J CO lines detected thanks to Herschel in NGC 7023 NW, the Orion Bar, and Carina have revealed a correlation between the thermal pressure and the intensity of the far-UV radiation field at PDR interfaces \citep{Joblin2018, Wu2018} globally in agreement with the theory \citep{Bertoldi1989, Bertoldi1996, Bron2018}. Additional studies have proposed similar relationships between thermal pressure and \rev{FUV} field strength using different tracers and across various interstellar environments, as discussed by \citet{Wolfire2022}. These findings indicate that our understanding of the interplay between thermal pressure and UV radiation in such regions remains incomplete.

The high thermal pressures that are found in dense PDR also result in the development of a new chemistry closer to the ionization front than previously anticipated because of the presence of far UV photons in the dense compressed gas. UV photons excite H$_2$ molecules in excited vibrational levels (they are pumped in Lyman and Werner bands followed by fluorescence then cascades in the ground electronic level as detailed in \citet{Black1987} allowing the endothermicity of some reactions to be overcome thanks to the internal energy of H$_2$ (e.g. C$^+$ + H$_2$ $\rightarrow$ CH$^+$ + H, O + H$_2$ $\rightarrow$ OH + H, see \citealt{Agundez2010, Zanchet2013, Faure2017, Goicoechea2021}), which then initiates chain of reactions leading to important observable tracers such as CO \citep{Joblin2018, Goicoechea2019} or CH$^+$ \citep{Zannese2025b}.
These processes highlight the complex interplay between \rev{FUV} radiation and the chemical and physical structure of PDRs. 

The Meudon PDR code \citep{LePetit2006} manages to model the complexity of PDRs by assuming a stationary, one-dimensional system and  focusing on the physico-chemical processes of all species. This allows for the extraction of the maximum amount of information from the observed emission lines in order to understand and constrain the different physical and micro physical processes at work in the PDRs.

While previous ground observations of PDRs in the near infrared had detected only a handful of ro-vibrational transitions of H$_2$, the IGRINS spectrometer has recently allowed the detection of tens of H$_2$ ro-vibrational lines in multiple galactic PDRs \citep{Kaplan2017,Le2017,Kaplan2021}  reaching vibrational emission lines that even the JWST has not detected. Indeed, the spectral sensitivity of IGRINS allows for the detection of $\text{H}_2$ lines with an excellent Signal-to-Noise Ratio ($\text{SNR}$) for vibrational levels ranging from $v=1$ to $v=13$, thereby complementing the spectral range of $\text{NIRSpec}$, which primarily detects pure rotational $\text{H}_2$ and ro-vibrational lines ranging from $v=1$ to $v=6$. The complementarity between the spectral resolution of $\text{IGRINS}$ coupled with the spatial resolution (and spectral sensitivity to a lesser extent) of the $\text{JWST}$ then enables the commencement of a new era in the study of $\text{PDRs}$ with high instrumental resolution.
In this paper, we use IGRINS observations of five classical PDRs (Horsehead, IC63, NGC2023, Orion Bar, and S140) providing more than one hundred H$_2$ ro-vibrational lines (from $v=1$ up to $v=13$). We use these lines to better constrain the physical conditions within PDRs and to gain deeper insight into the physical and chemical processes at work. In Section~\ref{Observations}, we present the observations and the reduction of the data. In Section~\ref{Method}, we explain how we compare these observations to the results of the Meudon PDR code to deduce the physical conditions in the PDRs. Results such as the deduced physical conditions in each PDRs are presented in Section~\ref{Results}. We discuss the results in Section~\ref{Discussion}. 

\section{Observations}\label{Observations}

\begin{table*}[h]
 \caption[]{\label{Table:PDRRef}PDR properties}
 \setlength{\tabcolsep}{3.5pt}
 \centering
\begin{tabular}{lcccccc}

 \hline \hline
PDR					&		S140 						&	 IC63 							& 	Horsehead 							& NGC2023		 					& 	Orion Bar \\

\hline
Type					& 	H  {\small II} region edge 				& Reflection nebula						& Dark nebula edge							&Reflection Nebula						& 	H  {\small II} region edge \\
$G_0$				&	 $1-4\times10^2\,$  \tablefootmark{a}  	& $58\tablefootmark{g}-150$\tablefootmark{h}	&  	 $100$ \tablefootmark{l}					& $10^3-10^4\,$ \tablefootmark{p} 			& 	$1-7.1\times10^4\,$ \tablefootmark{u}   \\
$P_{\text{th}}(Kcm^{-3})$	&	$2-10\times10^6\,$\tablefootmark{b}	& $0.4-3.5\times10^6\,$\tablefootmark{i} 	&	$1.3 - 9.2\times10^6\,$\tablefootmark{m}	& $3.5-35\times10^6\,$\tablefootmark{q}      			&	$1-3\times10^8\,$ \tablefootmark{v} \\
$d_{proj}$ to Star (pc)	& 	1.85  \tablefootmark{c}				& $1.3\tablefootmark{h}  $					&	3.5\tablefootmark{l}  					& 0.15\tablefootmark{r}				&	0.228 \tablefootmark{w} \\

Distance to Earth (pc)	& 	$764\pm27$\tablefootmark{d}			& $164\pm4$ \tablefootmark{j} 			& 	$387.5\pm 1.3$\tablefootmark{n} 			&399$\pm4$\tablefootmark{r}			&	$414\pm7$\tablefootmark{x} \\
\hline

llluminating Star 		& HD 211880 							& $\gamma$ Cas			  			& 	$\sigma$ Ori Aa, Ab and B	  			& HD 37903							& 	$\theta^1$ Ori C  \\

Spectral Type			& B0.5V\tablefootmark{a} 					&B0.5IV\tablefootmark{k} 					&	O9.5V\tablefootmark{o} 					& B1.5V\tablefootmark{s}					&	O7V\tablefootmark{y} \\

$T_{\textrm{eff}} (K)$	& 29,000\tablefootmark{a} 				&25,000\tablefootmark{k} 					&	33,000\tablefootmark{o} 					& 23,700\tablefootmark{s}					&	39,000\tablefootmark{y} \\

log (g)				&	(4\tablefootmark{e})					& 3.50 \tablefootmark{k} 					& 	4.20 	\tablefootmark{o} 					& (4\tablefootmark{t}) 					&	4.10\tablefootmark{y} \\

$M_\star$   			&  (18$M_\odot$\tablefootmark{e})			&  $13-18M_\odot$ \tablefootmark{k} 		&  	$20M_\odot$\tablefootmark{o}   			&  ($9.7M_\odot$\tablefootmark{t})  			&  	$45M_\odot$\tablefootmark{y}  \\

$R_\star$   			&    ($7R_\odot$\tablefootmark{f}) 			&   $10R_\odot$ \tablefootmark{k} 			&  	$5.6R_\sun$\tablefootmark{o} 				& ($5.8R_\odot$\tablefootmark{t})  			&   	$10R_\odot$\tablefootmark{y}   \\
\hline
\end{tabular}
\tablefoot{ We write in parentheses parameter values for which no direct estimate was found in the literature. The value is then our proposition based on indirect informations from the literature.\\
\tablefoottext{a} {\citet{Timmermann1996,Spaans1997,Wyrowski1997,Habart2004},}
\tablefoottext{b} {The thermal pressure is estimated as $P_{\text{th}}=n \times T$ using the rotational excitation temperature of H$_2$,} 
\tablefoottext{c} {Considering 7' difference from \citet{Poelman2005} with distance derived from \citet{Hirota2008}}, 
\tablefoottext{d} {Distance to the L1204 cloud associated to S140 from \citet{Hirota2008},}
\tablefoottext{e} {Values for a B0.5V star using the prescription of \citet{Kurliliene1981} on stellar parameters according to their spectral type,} 
\tablefoottext{f}  {Star radius deduced from the bolometric luminosity from \citet{Timmermann1996},}
\tablefoottext{g} {\citet{Eiermann2024},} 
\tablefoottext{h} {\citet{Andrews2018},} 
\tablefoottext{i} {$P_{\text{th}}=n \times T$ using n and T estimates from \citet{Thi2009} and \citet{Andrews2018},}  
\tablefoottext{j} {\citet{vanLeeuwen2007},} 
\tablefoottext{k} {\citet{Sigut2007},}
\tablefoottext{l} {\citet{Abergel2003},}  
\tablefoottext{m} {\citet{Habart2005,Habart2011,Hernandez2023},}
\tablefoottext{n} {\citet{Schaefer2016},}
\tablefoottext{o} {\citet{SimonDiaz2015},}
\tablefoottext{p} {\citet{Sheffer2011} for the upper limit, lower limit representing combination of new star distance from Gaia EDR3 \citep{GaiaCollaboration2021} and a possible difference between the true and projected star-PDR distances,} 
\tablefoottext{q} {\citet{Fleming2010,Sheffer2011},}
\tablefoottext{r} {Gaia EDR3 \citep{GaiaCollaboration2021},}
\tablefoottext{s} {\citet{Compiegne2008},}
\tablefoottext{t} {Values for a B1.5V star using the prescription of \citet{Kurliliene1981} on stellar parameters according to their spectral type,} 
\tablefoottext{u} {\citet{Marconi1998,Peeters2024},}
\tablefoottext{v} {\citet{Joblin2018,Goicoechea2016},} 
\tablefoottext{w} {\citet{Peeters2024},}
\tablefoottext{x} {\citet{Menten2007},} 
\tablefoottext{y} {\citet{SimonDiaz2006}.}
}
\end{table*}

\subsection{\rev{The Immersion Grating INfrared Spectrometer (IGRINS)}}

In this paper, we use H$_2$ emission line observed in five PDRs with the Immersion Grating INfrared Spectrometer (IGRINS) on the 2.7 m Harlan J. Smith Telescope at McDonald Observatory.  The data were initially presented in \citet{Kaplan2021} with normalized fluxes. For the present article, the data have been fully calibrated, resulting in the use of the absolute flux (see Section~\ref{Sec:data_reduction}).

IGRINS is a cross-dispersed scale spectrometer that observes the near-IR H and K bands ($1.45$--$2.45$~$\mu$m) simultaneously at a spectral resolving power of $R\sim45000$ or 7--8~km~s$^{-1}$  \citep{yuk2010, wang2010, gully2012, moon2012, han2012, park2014, oh2014, jeong2014, mace2016, mace2018, Sawczynec2025}. The large wavelength coverage and high resolving power of IGRINS allow us to avoid line blends and resolve up to hundreds of individual H$_2$ ro-vibrational transitions in each PDR. At McDonald Observatory, the IGRINS slit is $1 \times 15$ arcsec$^2$ on the sky. Each target PDR was observed at a single slit position (specified in \citealt{Kaplan2021}).

\subsection{The PDR sample}

\rev{The five studied PDRs are:} Sharpless 140 (S140), IC63, the Horsehead Nebula, NGC2023, and the Orion Bar. \rev{A detailed description of these PDRs can be found in Appendix~\ref{Appendix:PDRs}.} 

We characterize these regions using two main parameters: 
(1) the intensity of the FUV radiation field illuminating the PDR surface, expressed as $G_0$\footnote{We use 
$G_0 = \frac{1}{5.6 \times 10^{-14}\,\mathrm{erg\,cm}^{-3}} \int_{912\,\text{\AA}}^{2400\,\text{\AA}} u(\lambda)\, d\lambda$, 
as described in Appendix C of \citet{LePetit2006}.}. 
This value represents the ratio of the incident FUV energy density at the PDR surface to that of the Habing interstellar radiation field (ISRF) \citep{Habing1968}; 
and (2) the thermal pressure of the gas, P$_{th}$, \rev{given that previous studies have shown that constant-pressure models better reproduce the observations \citep{Marconi1998,Allers2005,Joblin2018,Wu2018}.} 

Table~\ref{Table:PDRRef} summarizes  previous estimations of the physical conditions in these regions from the literature. This sample includes three PDRs with previously estimated conditions corresponding to "low excitation” PDRs ($G_0$ values on the order of a few hundreds): Horsehead, IC63, and S140, and one with previously estimated conditions corresponding to "high excitation” PDRs ($G_0$ values larger than several thousands): the Orion Bar. NGC2023 seems to correspond to more intermediate conditions between the two categories.

\subsection{Data reduction}\label{Sec:data_reduction}

The data reduction follows what is described in \citet{Kaplan2017, Kaplan2021}, which we summarize here.   Cosmic rays are removed from the raw data using Version 0.4 of the Python implementation of LA-Cosmic cosmics.py\footnote{Python implementation of LA-Cosmic \citep{dokkum2001} by Malte Tewes: \url{https://web.archive.org/web/20180624102811/http://obswww.unige.ch/~tewes/cosmics_dot_py/}}.  The raw IGRINS data are reduced using v2.2.0 of the IGRINS Pipeline Package (PLP)\footnote{IGRINS Pipeline Package (PLP): \url{https://github.com/igrins/plp}} \citep{igrins-plp-2.2.0}.  The H and K bands are reduced separately.   The reduced IGRINS spectra are in the form of 2D arrays that are 100 pixels along the spatial axis and separated by echelle order.  For data analysis, we use the ``plotspec''\footnote{\rev{Plotspec: \url{https://zenodo.org/records/15595628}}} software which is designed to process 2D emission line data from IGRINS.  The echelle orders are stitched together to form one long 2D spectrum for each band. For every target except the Horsehead Nebula, the 2D spectra are continuum subtracted using a wide band running median. 
We apply a first order correction to the sky emission which mainly arises from rapidly varying hydroxyl (OH) transitions by taking the difference between two OFF position exposures and cross correlating the difference to the sky emission line residuals in amplitude and pixel shift (due to flexure). The best fit of this difference to the sky emission line residuals is then subtracted from the science spectrum.

For the data presented in this paper, we have made several improvements upon the data analysis that is described in \citet{Kaplan2017, Kaplan2021}, which include absolute flux calibration and updated line flux extraction.
Absolute flux calibration is an important step because the absolute line fluxes scale to absolute rovibrational level column densities of excited H$_2$ molecules.  This allows a direct comparison between the observed empirical H$_2$ line fluxes and the line fluxes predicted by the Meudon PDR code models, and provides important constraints to the models. Each PDR target had an A0V standard star observed at a similar airmass. 

 Absolute flux calibration requires good estimates of two factors about the standard star spectrum. The first aspect we need to estimate is the fraction of light from the star that passes through the slit. The standard star was observed by nodding between two positions along the slit. The IGRINS PLP gives the star's point spread function (PSF) along the spatial axis of the slit.  We fit a Moffat function to the standard star's PSF at these two slit positions, and project these fits onto a 2D array representing the sky assuming the PSFs are radially symmetric and smeared out  in the east-west direction by 1.5 arcsec to account for error in the telescope guiding. We then apply a mask to the 2D array that is the size  and position angle of the slit to estimate the fraction of light from the standard star that passed through the slit.  We do this separately for the H and K bands.  The wavelength dependent fraction of light through the slit is estimated by fitting a linear function to the fraction  estimated for the H and K bands vs. the inverse each band's central wavelength.  Properly accounting for this wavelength dependence is important in order to get accurate estimates of the dust extinction.

The second element we need to estimate for absolute flux calibration is the intrinsic spectrum of the standard star. We do this by using the Gollum\footnote{Gollum: \url{https://github.com/BrownDwarf/gollum/}} software \citep{Shankar2024} to read in and match Phoenix model stellar atmosphere synthetic spectra  \citep{husser2013}  to the standard stars.  The Phoenix model grid cover a wide range of effective temperatures, metallicities and surface gravities.  We vary these stellar parameters, along with the radial and rotational velocity, until we find the model that best fits the H and K band magnitudes and H I line profile shapes of the standard star.  In practice, we typically interpolate between two or more of the Phoenix synthetic stellar spectra to best match intermediate values of the standard star's stellar parameters that do not fall directly on the model grid. 
Once we have our estimates 
for the standard star's intrinsic spectrum and fraction of through the slit and its intrinsic spectrum, and knowing the ratio of exposure time on the target PDR to standard star, we can absolute flux calibrate and telluric correct our target PDR spectrum.  

We then extract the H$_2$ line fluxes and calculate extinction corrections following the same procedures described in \citet{Kaplan2021}, but in this paper we have calculated new extinction values 
since we have made major updates to our flux calibrations. 
As in \citet{Kaplan2021}, we calculate our new extinction corrections from pairs of H$_2$ transitions with the same upper ro-vibrational level. The extinction correction was compared to that derived from JWST observations \citet{Peeters2024,Zannese2025,Abergel2024} along the same lines of sight, and the one proposed in this study is found to be equivalent. We exclude line pairs that are close together in wavelength, since they do not well sample the differential extinction.
Our new estimates for total K-band extinction ($\mathrm{A}_\mathrm{K}$) in the PDRs are 0.87, 0.55, 1.5, 0.51 and 0.73 for S140, IC63, the Horsehead Nebula, NGC2023 and the Orion Bar.  These estimates of $\mathrm{A}_\mathrm{K}$ are slightly larger than the ones from \citet{Kaplan2021}, with the Horsehead Nebula showing the largest difference ($A_K \sim 1.5$ in this paper vs. 0.70 in \citealt{Kaplan2021}).  This large difference is because our observation of the Horsehead Nebula's standard star showed a large wavelength dependence in the fraction of light from the star passing through the slit.  This highlights the importance of properly accounting for the standard star slit throughput in our flux calibrations, otherwise our extinction estimate could be incorrect. In the remainder of this article, we only use the extinction-corrected H$_2$ line fluxes.
	
\section{Method}\label{Method}
\subsection{Meudon PDR code}\label{model_description}

We use the latest version of the Meudon PDR code\footnote{Meudon PDR code: \url{https://ism.obspm.fr/pdr.html}} \citep{LePetit2006}, version 7 released in 2024, to model each PDR and to best reproduce the H$_2$ lines observed by IGRINS. The model assumes a stationary plane-parallel slab of gas and dust illuminated by a UV radiation field. In this paper, the slab of gas is assumed to have constant thermal pressure, which has been shown to better explain the density stratification in dense PDRs \citep{Marconi1998,Habart2005,Joblin2018}. 

At each point in the cloud, the code iteratively solves:
\begin{enumerate}
\item the radiative transfer, from the FUV to the radio domain, to determine the penetration of the incident radiation field. The code accounts for continuum absorption by dust and by the ionization of species such as carbon, non-isotropic scattering by dust, as well as absorption in UV lines by key species such as H and H$_2$ \citep{Goicoechea2007}. For H$_2$, all UV transitions from the ground electronic state X to the B, B$'$, C$^-$, C$^+$, D$^-$, and D$^+$ electronic levels are considered (36,391 lines in total)\footnote{To reduce computing time, weak lines that do not affect shielding significantly are treated using the \citet{Federman1979} approximation.}.This formalism is also applied to CO, $^{13}$CO, and C$^{18}$O. This detailed treatment, coupled with the calculation of H$_2$ level populations, allows a precise, level-by-level determination of H$_2$ photodissociation rates. \revA{However, the present models do not include the photoionization of vibrationally exited H$_2$ \citep{Ford1975} or the v-dependent charge transfer between H$_2$* and H$^+$ implemented in another version of the Meudon PDR code by \cite{Goicoechea2025} but found to play a minor role}.
\item the thermal balance, considering heating processes (photoelectric effect on dust grains, cosmic ray heating, exothermic chemical reactions,...) and cooling via emission lines from dozens of ions, atoms, and molecules, including C$^+$, C, O, CO and its isotopologues $^{13}$CO and C$^{18}$O. Depending on the physical conditions, H$_2$ can either cool or heat the gas. In dense UV-illuminated regions, H$_2$ heating contribution, coming from UV-pumping and fluorescence to ro-vibrational levels of the ground state followed by collisional de-excitations, can overcome its cooling contribution from collisionaly-excited pure rotational emission. Gas-grain collisions can also be a cooling or heating term. The computation of line emission cooling rates employs the formalism described in \citet{Gonzalez2008}.
\item the chemistry, using a network of 223 species linked by 3300 reactions. For H$_2$, the code simulates the formation on a MRN grain size distribution \citep{Mathis1977}, including both Eley-Rideal and Langmuir-Hinshelwood mechanisms \citep{LeBourlot2012}.
\item the statistical balance of quantum level populations for dozens of species. 
We do not assume Local Thermodynamic Equilibrium (LTE), and these populations are calculated considering radiative and collisional transitions, as well as level-specific chemical formation and destruction rates.
 For H$_2$, we include collisions with H$^+$, H, ortho- and para-H$_2$, and electrons with references detailed below. 
Different prescriptions are employed for the level-distribution of H$_2$ formation depending on the reaction pathway, with the most critical being the excitation of nascent H$_2$ upon formation on grains, that is discussed in Section~\ref{Section:Sizun}. We assume that H$_2$ destruction occurs with equal probability across all levels, except for two types of reactions:
For photodissociation, the code explicitly computes, level by level, the UV pumping in the Lyman and Werner bands followed by dissociation.
For reactions with level-specific reaction rates available, such as C$^+$ + H$_2$ \citep{Zanchet2013,HerraezAguilar2014}, O + H$_2$ \citep{Veselinova2021} or S$^+$ + H$_2$ \rev{(\citealt{Zanchet2013b} for $v=0$ and 1 and \citealt{Zanchet2019} for higher $v$)}, these are also taken into account to compute level-specific destruction rates for the detailed balance computation of H$_2$ level populations.
We also include ortho-to-para conversion of H$_2$ on dust grains \citep{Fukutani2013, LeBourlot2000, Bron2016}. This mechanism tends to thermalize the ortho-to-para ratio toward the grain temperature. By default, the Meudon PDR code assumes a 100\% conversion efficiency.
\end{enumerate}
The determination of level populations, chemical abundances and thermal balance is tightly coupled to radiative transfer, which also couples spatial positions together. Global iterations over the full spatial grid are thus also performed to allow the result of the local computations to affect other spatial positions.

In version 7 of the Meudon PDR code, many atomic and molecular data, as well as chemical reaction rates have been updated, in particular concerning H$_2$.
For H$_2$ collision rates with ortho and para H$_2$ we use \citealt{Flower1998b, Flower1999} supplemented by \citet{Wan2018} that provide rates up to the $v = 12$ and $J = 10$ level. For collisions with H, we implemented the rates by \citet{Lique2015} up to $v = 3$, $J = 8$, complemented by semi-classical rates by \citet{Bossion2018} for higher levels, up $v = 12$ and $J = 10$. For collisions with H$^+$ and e$^-$, data are from \citet{Gonzalez2021} and \citet{England1988}.
 
Once the code has converged, it provides the species densities, level populations, and gas and grain temperatures at each position, along with the emergent spectrum, including line intensities. This enables direct comparison with the observed line intensities. 

In this study, for each PDR, we investigate which combination of thermal pressure, $P_{\text{th}}$, and incident \rev{FUV} field intensity \revA{at the ionization front}, characterized by the $G_0$ parameter, best reproduces the IGRINS observations. The other input parameters of the Meudon PDR code are kept fixed (see Table~\ref{Table:Parameters}), and the elemental abundances adopt the standard values commonly used in previous studies \citep{Joblin2018}. 

\rev{Observational determinations of extinction properties inside the PDR layers are missing for most PDRs in our sample, due to the difficulty of measuring local extinction properties rather than global line-of-sight properties. For consistency, we decided to use a single extinction curve for modeling internal dust extinction in the five PDRs in this sample, corresponding to dense molecular gas and previously used in PDR models in \citet{Joblin2018} ($R_V = 5.62$, $N_\mathrm{H} / E(B-V) = 1.05\times10^{22}$ cm$^{-2}$ mag$^{-1}$, using a \citealt{Fitzpatrick1990} fit of the line of sight towards HD 38087 as representative of this type of environment). The $R_V$ value we use is very similar to the one ($5.5$) used by the PDRs4all consortium for specific studies of the Orion Bar PDR \citep{Peeters2024,VanDePutte2024,Habart2024}. The question of extinction of the incident FUV radiation field is further discussed in Sect.~\ref{extinction_discussion}}
 
\begin{table*}
 \centering
\setlength{\tabcolsep}{2pt}
 \caption[]{\label{Table:Parameters}Fixed parameters used for all PDR modeling of this paper.}
\begin{tabular}{lcccc}
\hline \hline
 Symbol 					&	Value 			&	Name						 & 		Note 	\\
 \hline
$A_\mathrm{V}^{tot}$  		&	30 mag  			&	  Size of the cloud in visual extinction &  		Chosen to ensure the inclusion \\
& & & of all main processes \\
\hline
Extinction Curve			& HD38087  			& Typical molecular line of sight			& \citet{Fitzpatrick1990}	 \\
$R_V$ 					& 5.62 				&   Total-to-selective extinction ratio		& \citet{Joblin2018}		 \\ 
$N_H/E(B-V)$ 				& $1.05 \times 10^{22}$ 	&  	Hydrogen column density to color excess ratio	& \citet{Joblin2018}		  \\
$r_d$	  				& $3\times10^{-7}$  to $3\times10^{-5}$ cm & Dust distribution (MRN distribution) in grain radius & 	 \\
\hline
$\zeta$    					& $5.0 \times 10^{-17} \, \mathrm{s}^{-1}$  		& Cosmic ray ionisation rate 			&  						 \\
$V_{turb}$     				& $2 \, \mathrm{km} \, \mathrm{s}^{-1}$ & Turbulent velocity (Doppler broadening) &   						\\
 Z 						& 1 			& Metallicity 						&  	Milky Way					\\

\hline
\end{tabular}
\end{table*}
 

 \subsection{FUV illumination and stellar spectra}\label{incident_UV_fields}
 
Most PDR modeling studies have assumed an isotropic illumination of the PDR, using the Interstellar Standard Radiation Field (ISRF) from \citet{Mathis1983} or \citet{Draine1978} multiplied by a scaling factor to simulate illumination by nearby O/B stars.
However, a sensitivity analysis has been performed in Appendix~\ref{Spectrum}, showing the importance of a more realistic modeling involving: (1) a beamed illumination of the PDR, (2) a more realistic spectrum shape (realistic stellar spectrum or at least a blackbody at the star's effective temperature).

We thus used a beamed illumination in all our models, with a specific stellar spectrum for each modeled PDR.
We used theoretical stellar spectra from the Pollux Database \citep{Palacios2010} based on stellar parameters (effective stellar temperature $T_{\textrm{eff}}$, mass of the star $M_\star$, and $\log(g)=\log(\frac{GM_\star}{R_\star^2})$) from the literature (cf. Table~\ref{Table:PDRRef}).
The stellar models selected are the closest in stellar parameter values found in the Pollux database, and are detailed in Table~\ref{Table:Stars}.
The ultraviolet radiation field incident on the PDR is then calculated by applying a spherical dilution factor based on Pollux model's stellar radius and the assumed star-PDR distance. \rev{We neglect dust extinction of the incident FUV stellar spectrum occurring within the ionized region that precedes the PDR (cf. Sect.~\ref{extinction_discussion} for a discussion of this assumption).} The results will be presented in terms of the resulting $G_0$ at the ionization front to facilitate comparison with existing literature.

The majority of stars listed in Table~\ref{Table:PDRRef} have been extensively studied spectroscopically, allowing for relatively precise constraints (such as $\gamma$ Cas or $\sigma$ Ori Aa). However, other stars like HD 211880 (for S140) or HD 37903 (for NGC2023) have received less attention in the literature, hindering our ability to identify a suitable stellar analogue. Given the spectral type of these stars, we derive stellar parameters from \citet{Kurliliene1981}. 
A sensitivity analysis was performed using different stars from Pollux with the stellar effective temperature. The results indicate that variations in stellar mass have a minimal impact on the spectrum between 912 and 2400 \AA, and consequently, cause very little change in $G_0$.

\begin{table}[h]
 \centering

\setlength{\tabcolsep}{3.5pt}
 \caption[]{\label{Table:Stars}Stellar models selected for each PDR }
\begin{tabular}{lcccccc}
\hline \hline
PDR 		                 &	S140 		& 	IC63 		& 	Horsehead 		& NGC  2023 			 	& 	Orion Bar \\

\hline
$T_{\textrm{eff}}$ [K]		& 29,000			&25,000			&	33,000			& 23,000					&	39,000\\
log(g)		                 &4.20			&3.80			&4.26				&	4.00					&4.24\\
$R_\star$		                  &$7R_\odot$		&$9R_\odot$		&$5.6R_\odot$			&   5.8R$_\odot$			&$10R_\odot$\\
\hline
\rev{$H^{\mathrm{LW}}$  }		& 			0.29	& 		0.25		&			0.32		&	0.22		&		0.32		\\
\hline
\end{tabular}
\tablefoot{\revAA{Spectra originating from the Pollux DataBase \citep{Palacios2010} for each PDR, based on the observational parameters cited in Table~\ref{Table:PDRRef}.} \rev{$H^{\mathrm{LW}}$  is the hardness of the FUV field defined in Eq.~\ref{eq:Hardness}}}
\end{table}

\rev{As the spectral shape of the incident field becomes strongly different between the stellar types used here and the scaled ISRF often used in PDR modeling, the $G_0$ parameter becomes insufficient to fully describe the incident radiation field. We propose to add a second parameter describing the hardness of the FUV field,
\begin{equation} \label{eq:Hardness}
H^{\mathrm{LW}} = \frac{\int_{912\,\text{\AA}}^{1108\,\text{\AA}} u(\lambda)\, d\lambda}{\int_{912\,\text{\AA}}^{2400\,\text{\AA}} u(\lambda)\, d\lambda} ,
\end{equation}
giving the fraction of $G_0$ located in the 912 to 1108 \AA\ band. As demonstrated Appendix~\ref{Spectrum}, the specific wavelength range between 912 and 1108 \AA\ is of critical importance for H$_2$ absorption lines (compared to the 912--2400 \AA\ band used in the $G_0$ definition) as it corresponds to the Lyman-Werner band transitions responsible for H$_2$ pumping and dissociation (from $v=0$, low $J$ levels). For the standard Mathis ISRF, $H^{\mathrm{LW}}$ is approximately $0.15$. In contrast, the irradiating fields in our modeled sources exhibit a significantly harder spectrum, with $H^{\mathrm{LW}}$ reaching $0.22$ for NGC 2023, and up to $0.32$ for the Orion Bar.}

\subsection{PDR model grids}
For each object, we create a grid of PDR models. We create one model grid for each of the studied PDRs, each employing a specific stellar spectrum 
The two parameters we explore are the thermal pressure in the PDR, $P_{\text{th}}$ and the intensity of the incident radiation field, $G_0$ (this is done by varying the distance between the star and the PDR). 
The grids cover values of $G_0$ from 10  to $10^5$ and $P_{\text{th}}$ from $5\times10^5$ to $5\times10^9$~K~cm$^{-3}$. 

To account for geometrical effects when comparing our plane-parallel models to observations of a corrugated PDR surface, possibly tilted with respect to the line of sight, we introduce, as an additional free parameter, a geometrical scaling factor $\Omega$, by which we scale the face-on line intensities predicted by the PDR models, following \citet{Sheffer2011}.
This multiplicative factor accounts for several observational effects (as H$_2$ lines are optically thin):
inclination of the PDR surface with respect to the line-of-sight and with respect to the irradiation direction, beam dilution, line-of-sight intersecting the PDR surface multiple times. We assume these effects to be of order unity and thus restrain this free parameter between 0.1 and 10.
\subsubsection{Parameter adjustment through reduced $\chi^2$ minimization}

As a first approach to fitting our model grids to the observations, we use $\chi^2$ minimization.
Observational noise is assumed to be additive gaussian. 
To account for model uncertainties and discrepancies arising from comparing plane parallel models to real PDRs with a complex geometry (beyond what can be captured by the linear scaling factor $\Omega$), we introduce modeling uncertainties as a multiplicative lognormal distribution with $\sigma_\mathrm{mod}$ set so that $e^{\sigma_\mathrm{mod}}=1.3$, i.e. taking 30\% uncertainty as a reasonable guess of these hard-to-quantify uncertainties. 
As an approximation to the combination of these two uncertainties, we only account for the largest of the two uncertainties for each observed line. In our case, the observational relative error was consistently lower than 30\% for all observations and all spectral lines. 

Our reduced chi-squared ($\chi^2_R$) is then defined by:
\begin{equation}\label{eq:chi2}
  	  \chi^2_R=\frac{1}{N-d}\sum_{n=1}^N\frac{ \left[\log(I^{obs}_n) - \log\left(\Omega \times I^{mod}_n(P_{\text{th}},G_0) \right)\right]^2}{\sigma_\mathrm{mod}^2}
\end{equation}
Where $N$ is the number of observed H$_2$ lines, $d$ the number of free parameters, $I^{obs}_n$ the observed integrated intensity of line $n$, $I^{mod}_n(P_{\text{th}},G_0)$ the model integrated intensity of line $n$ for the specific parameter values $(P_{\text{th}}, G_0)$, and $\sigma_\mathrm{mod}$ the modeling uncertainty discussed above.

To estimate $I^{mod}_n(P_{\text{th}},G_0)$ between grid points, we interpolate between the H$_2$ line fluxes from the grid models with the radial basis function (RBF) method (using the Scipy\footnote{\url{https://scipy.org/}} implementation). By minimizing the $\chi^2_R$ across the continuous model grid, we identify the best interpolated model that most closely reproduces the observed line intensities. A $\chi^2_R$ close to 1 suggests a good agreement between the model and the data.

\subsubsection{Bayesian inversion with \textsc{Beetroots}}

To obtain error bars on the estimated physical parameters and to ensure robust results through the use of two distinct methodologies, we also employed the Bayesian inversion method presented in \citet{Palud2025} and implemented in the python package \textsc{Beetroots}\footnote{\url{https://github.com/pierrePalud/beetroots}}. 
\textsc{Beetroots} employs an advanced Markov Chain Monte Carlo (MCMC) sampler. This approach yields robust estimations and quantifies associated uncertainties. 

To mitigate the high computation cost of MCMC sampling, we trained an artificial neural network (ANN) to emulate the results of the grids of models as in \citet{Palud2025}, and use this emulator for sampling. We resorted to \texttt{NNBMA}\footnote{https://github.com/einigl/ism-model-nn-approximation} \citep{Palud2023} to build an ANN with a dense architecture, a 3rd order polynomial expansion and 8 hidden layers, using the GELU activation function. As we use 5 different grids, one per observed PDR, we trained 5 different ANNs. Each ANN was trained for 5 million epochs, achieving a mean absolute relative error of 2\% on the test set, reaching a level negligible compared to the observational uncertainties discussed above.

Subsequently, we used \textsc{Beetroots} to perform parameter inference for the five observed PDRs (using the corresponding five trained ANNs).
The posterior distribution thus obtained allows for a comprehensive characterization of the uncertainties and degeneracies in the parameter estimates.
While this posterior distribution is proportional to the likelihood as we use uniform priors on the parameter, \textsc{Beetroots} provides the mean of the posterior distribution (minimum mean square estimator) as final parameter estimates, which is in general not the same as its maximum, which $\chi^2$ minimization (equivalent to maximum likelihood) will yield.
 
\section{Results}\label{Results}
\subsection{Best Fitting models}

\subsubsection{Unconstrained fitting results}\label{sect:unconstrained_fits}

Using every dedicated model grid and the Bayesian inversion method, we determined the best-fit model for each PDR without any constraints on the parameters. 

\begin{itemize}
\item For S140, we found $G_0 = 504$, $P_{\text{th}} = 9.5 \times 10^5$ K cm$^{-3}$, and $\Omega = 9.8$. \rev{The corresponding reduced chi-squared, $\chi^2_R$, is 1.7 and 95\% of the lines are reproduced within a factor 2.}
\item For IC63, the best-fit parameters are $G_0 = 542$, $P_{\text{th}} = 2.5 \times 10^6$ K cm$^{-3}$, and $\Omega = 3.5$ \rev{with a $\chi^2_R$ = 1.7 and also 95\% of the lines reproduced within a factor 2.}
\item For he Horsehead Nebula is best modeled with $G_0 = 292$, $P_{\text{th}} = 1.2 \times 10^6$ K cm$^{-3}$, and $\Omega = 9.7$ \rev{with a $\chi^2_R$ = 2.1 and 94\% of the lines reproduced within a factor 2.} 
\item For NGC2023, we obtained $G_0 = 800$, $P_{\text{th}} = 4.1 \times 10^6$ K cm$^{-3}$, and $\Omega = 6.2$, \rev{with a $\chi^2_R$ = 1.8 and 93\% of the lines reproduced within a factor 2.}
\item For the Orion Bar, the best-fit parameters are $G_0 = 1,502$, $P_{\text{th}} = 6.5 \times 10^7$ K cm$^{-3}$, and $\Omega = 6.8$ \rev{with a $\chi^2_R$ = 1.25 and 92\% of the lines reproduced within a factor 2.} 
\end{itemize}

\begin{table*}[h!]
 \centering
\setlength{\tabcolsep}{8pt}

 \caption[]{\label{Table:BestFit}Most relevant models for each PDR }
\begin{tabular}{lcccccc}
\hline \hline
\textbf{PDR} 					&	S140 					& 		IC63 			& 	Horsehead 		& NGC2023 				 & 		Orion Bar \\
\hline

$G_0$	\tablefootmark{a} 						& 	$504$					& 	$542$				& 	$258$			& 1,018 			&		10,280		\\
(equivalent d in pc)\tablefootmark{b} 				&	(\textit{0.98})				&	(\textit{0.86})			&	(\textit{2.4})			& (\textit{0.36})			&	 (\textit{0.44}) \\

68\% CI on $G_0$ 					& 	$430-580$				& 	$470-620$			& 	$225-290$		& 990 - 1,050			& 	10,000 - 10,500		\\
\hline

$P_{\text{th}} [\text({K} \, \text{cm}^{-3}]$				& 	$9.5\times10^5$			& 	$2.5\times10^6$		& 	$6.3\times10^6$		& $5\times10^7$			&	 $5.2\times10^7$\\
68\% CI  on $P_{\text{th}}$				& $8.3\times10^5-1.1\times10^6$	& 	$2-3.2\times10^6$		& 	$5.4-7.2\times10^6$ 		& $4.9 - 5.1\times10^7$ 					&	$4.9-5.6\times10^7$\\
\hline

$\Omega$						&	9.8						&	2.5					&	8.1					& 9.9							& 4.2	\\
 68\% CI on $\Omega$			&	9.5 - 10					&	3 - 4					& 7.1 - 9					& 9.8 - 10 						& 4.1 - 4.5	\\
\hline
$\chi_R^2$					&	1.7 						& 	1.7 					& 	2.5 					&3.6				&	 2.26\\
Lines within a factor of 2			&	95\% 					& 	95\%					& 	92\%					& 85\% 			&	 86\% \\
\hline
\end{tabular}
\tablefoot{\revAA{\rev{obtained} from Bayesian inversion with additional constraints for the Horsehead Nebula, NGC2023 and Orion Bar. Additional metrics as 68\% CI (from \textsc{Beetroots}),  $\chi^2$ minimization for $\chi_R^2$ estimation and the \% of lines reproduced within a factor of 2 are provided. Each PDR is modeled with a specific illuminating stellar spectrum.}\\
\rev{The models for S140 and IC63 presented in the table correspond to the unconstrained models presented in Section \ref{sect:unconstrained_fits}. For the Horsehead Nebula, NGC2023 and Orion Bar, the constrained models differ from the unconstrained models presented in Section \ref{sect:unconstrained_fits}}\\
\rev{(a)The $G_0$ is calculated using the specific stellar spectrum for each PDR (see Table\ref{Table:Stars})}.\\
(b)The distance between the star and the PDR is deduced from the values of $G_0$.}

\end{table*}

\begin{figure*}[htbp!]
    \centering
    \subfloat[S140]{\includegraphics[width=0.47\textwidth]{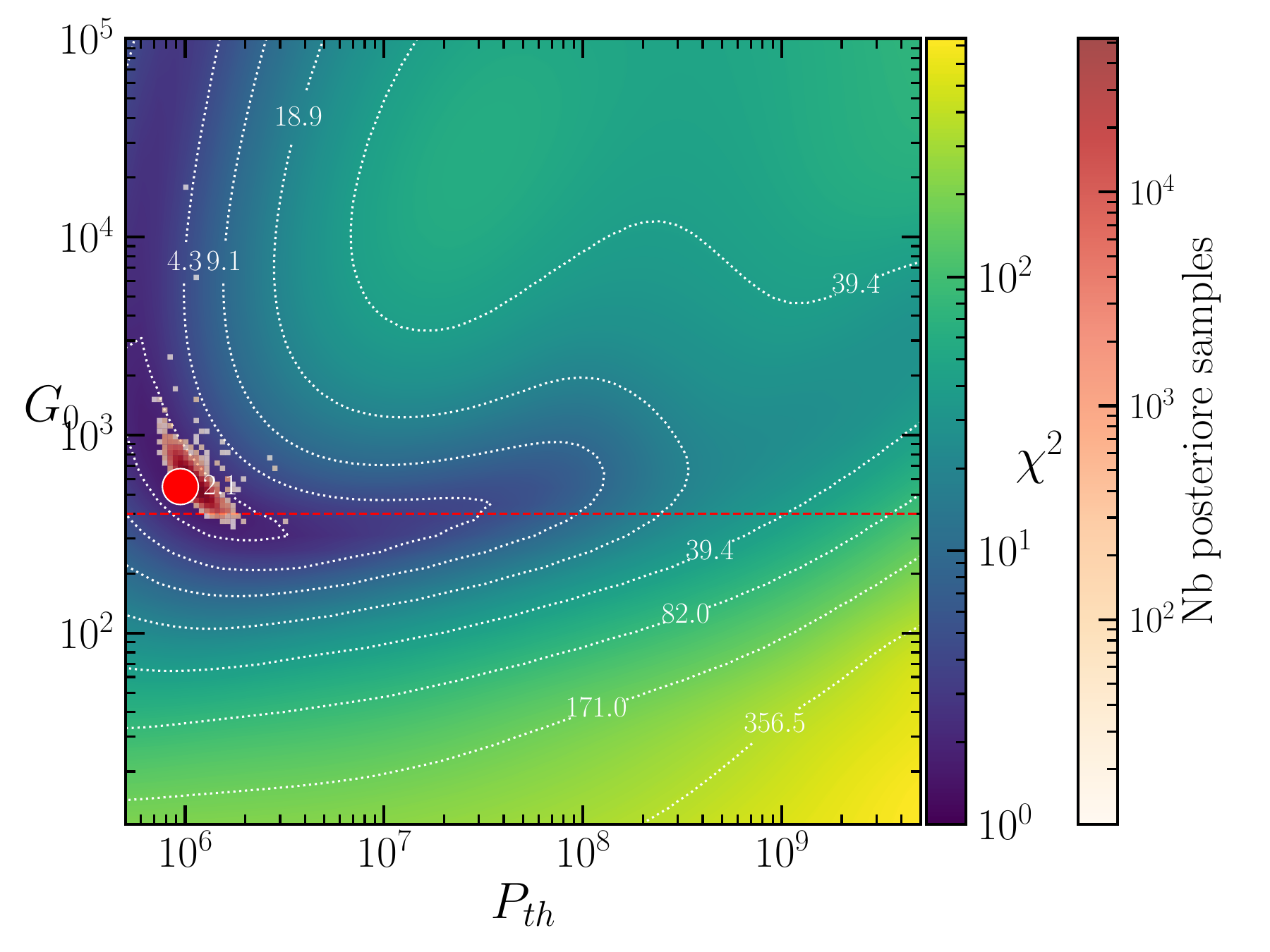}}
    \subfloat[IC63]{\includegraphics[width=0.47\textwidth]{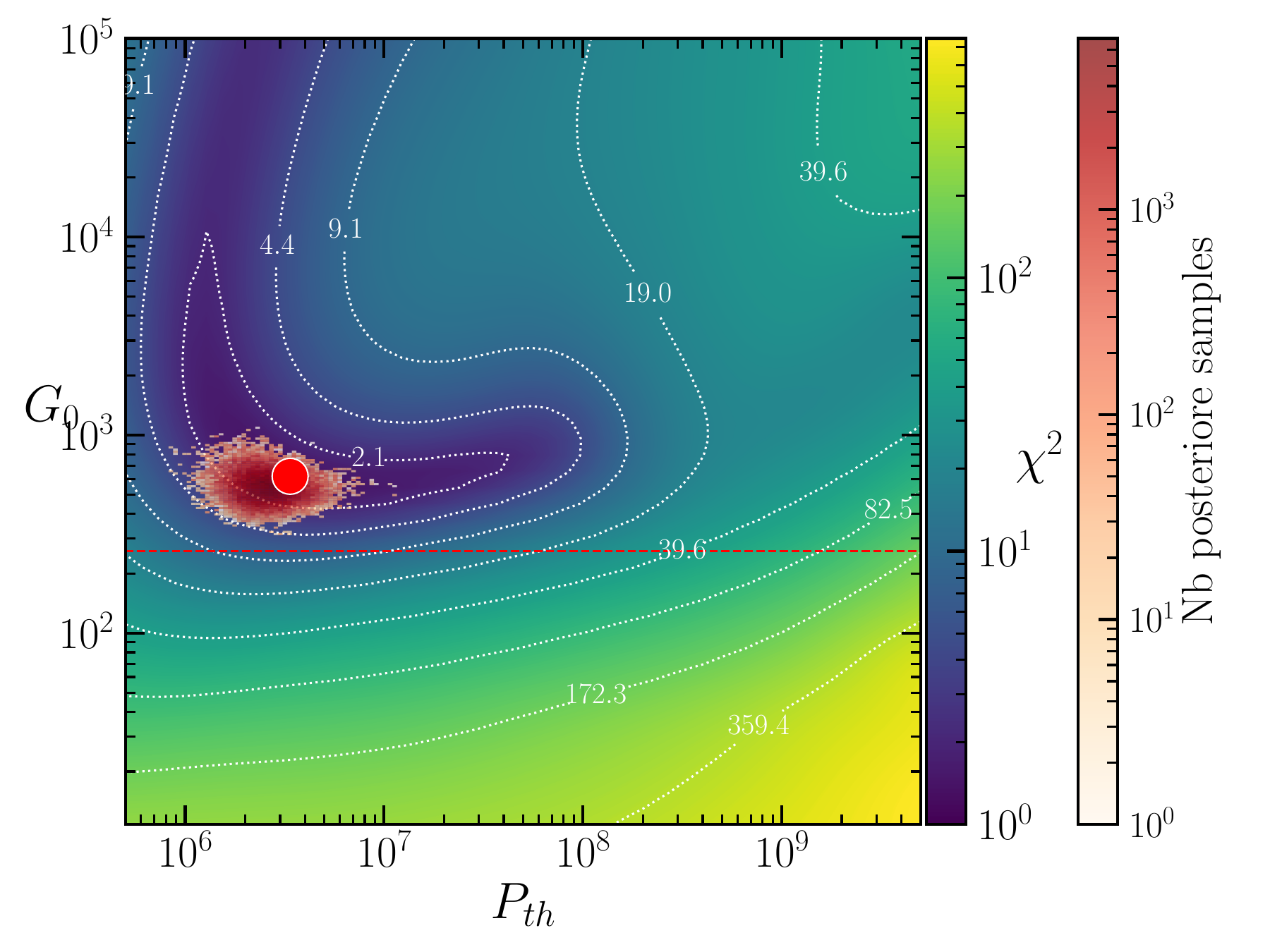}}
   
    \subfloat[Horsehead Nebula]{\includegraphics[width=0.47\textwidth]{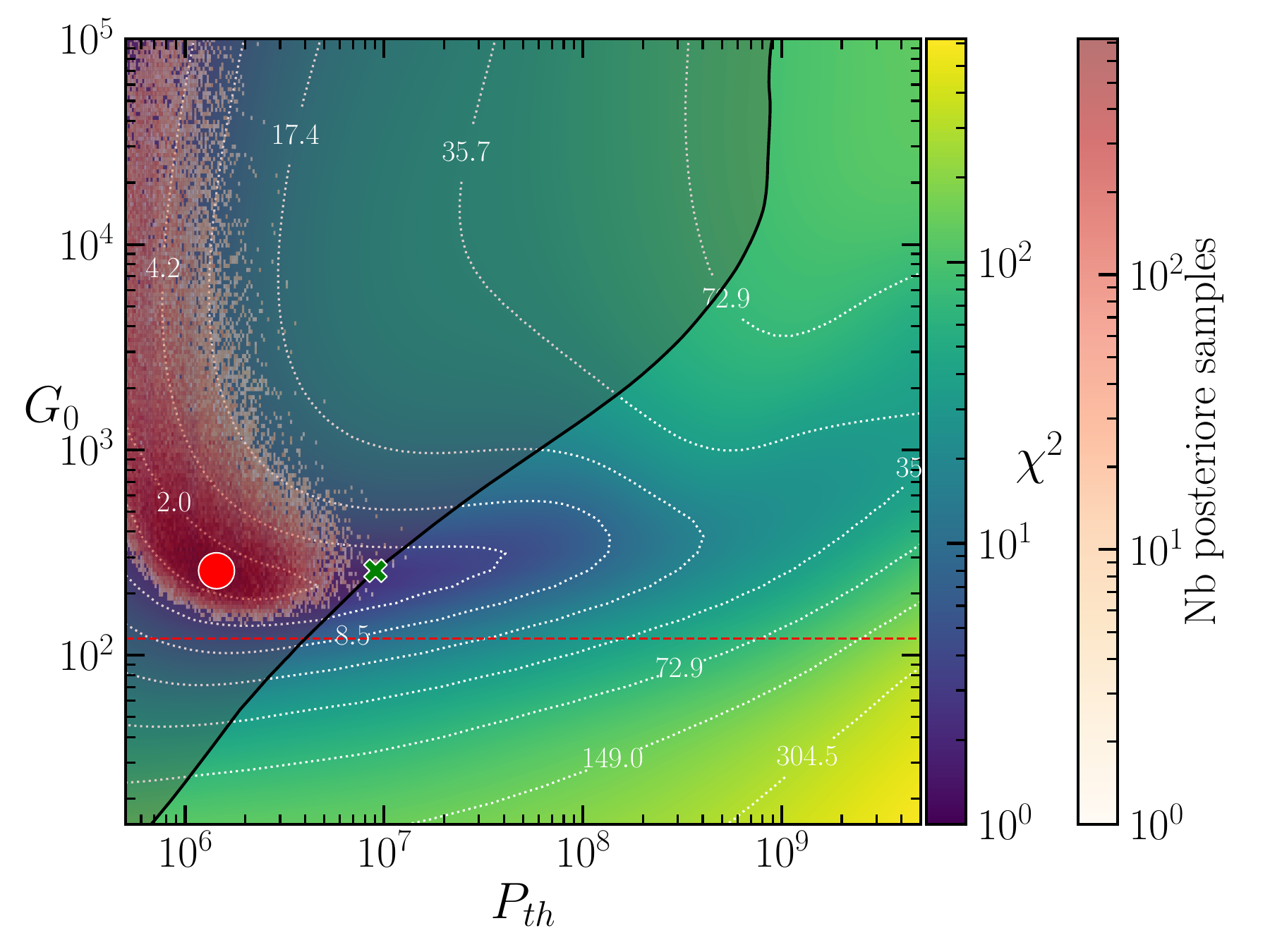}}
    \subfloat[NGC2023]{\includegraphics[width=0.47\textwidth]{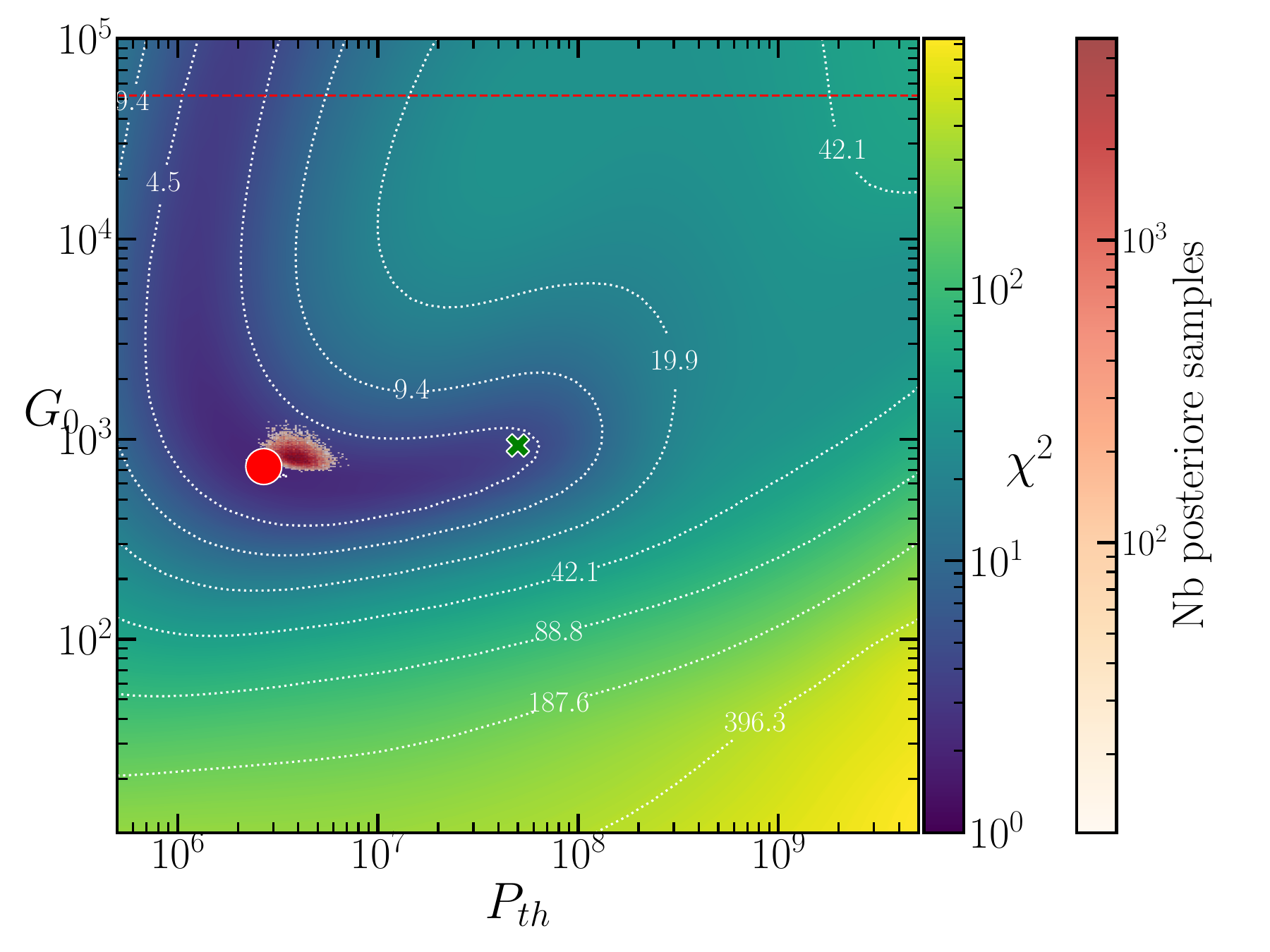}}

    \subfloat[Orion Bar]{\includegraphics[width=0.45\textwidth]{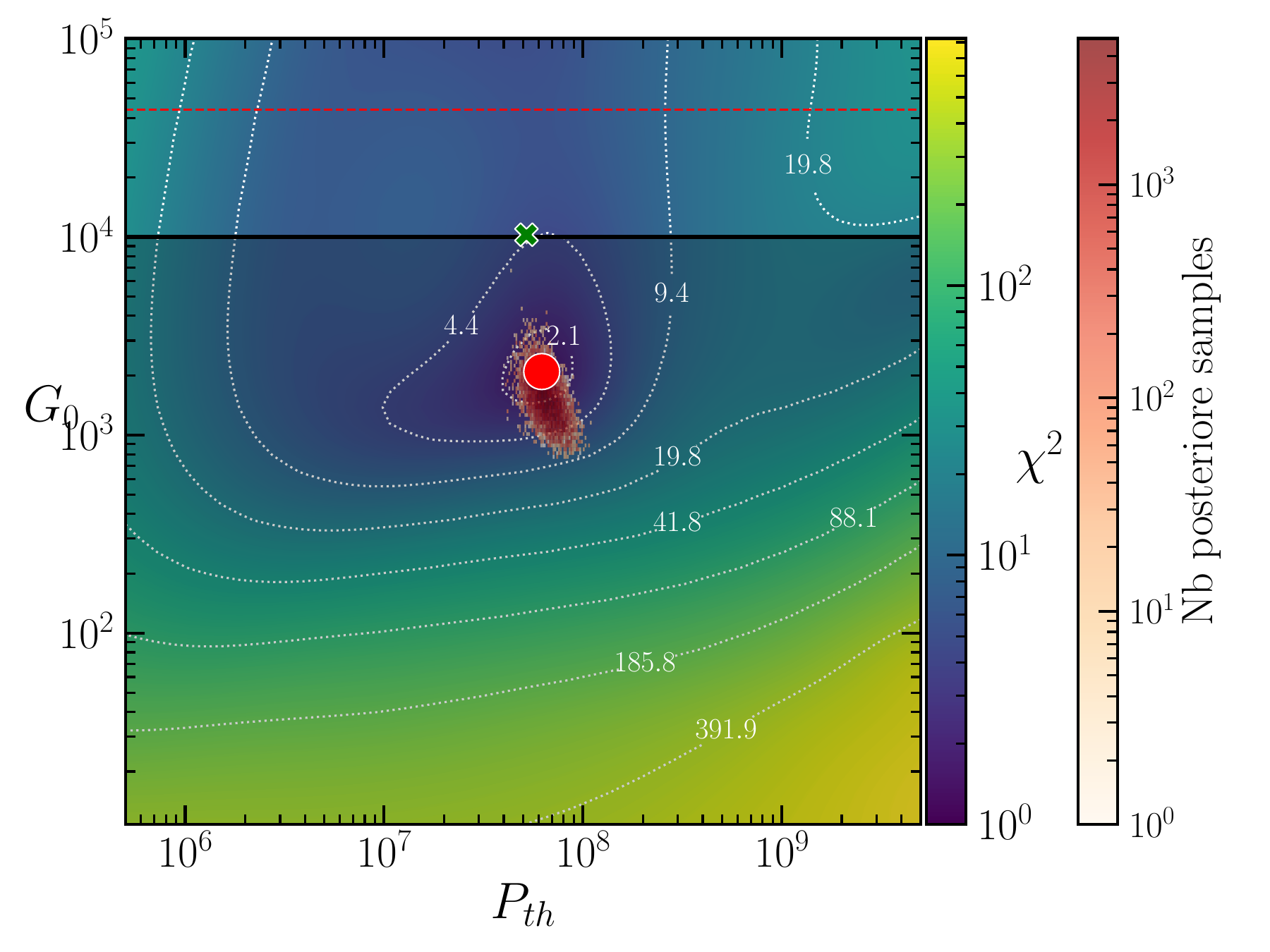}}
    \caption{$\chi^2_R$ contour map (blue-green color bar) in the $P_\mathrm{th}-G_0$ plane (at fixed $\Omega$ value corresponding to the minimum-$\chi^2_R$ model), and 2D histogram of the posterior samples generated with the MCMC method implemented in \textsc{Beetroots} (white-red color bar), for each PDR. The red point represents the best-fit model found with Bayesian inversion method.  Additional constraints are shown as black lines with darkened regions which are removed from the explored parameter space: For the Horsehead nebula, upper limit on the distance of the H/H$_2$ transition to the ionization front; 
    for NGC2023, a prior of P$_{\text{th}} \geq 5\times10^7\, \text{K}\, \text{cm}^{-3}$ was added in the posterior exploration; for the Orion Bar, lower limit $G_0\geq10^4$. The best proposed model accounting for the additional constraints is represented with a green cross. The red dashed line corresponds to the FUV irradiation taking the star-PDR distance to be equal to the plane-of-the-sky projected distance when available.}
    \label{fig:contour}
\end{figure*}

Figure~\ref{fig:contour} shows for the five studied PDRs the $\chi^2_R$ in the 2-parameter plane of $G_0$ and $P_{\text{th}}$, with $\Omega$ fixed at the value of the minimum-$\chi^2_R$ model (blue-to-yellow color map), as well as the 2D histogram of the posterior samples obtained with \textsc{Beetroots} (white-to-red color map). 
We observe a strong agreement between the two methods, with the 2D histogram closely overlapping the contour from the $\chi^2_R$ \rev{and the best fit estimate agreeing within a factor of 1.2}. Minor shape differences, particularly for the Orion Bar, are expected as the $\chi^2$ map represents a cut of the parameter space at a single $\Omega$ value, while the MCMC histogram integrates over all $\Omega$ values.

Red dots indicate the best-fit models from the Bayesian inversion method without any additional constraints.
For the three low-excitation PDRs (S140, IC63 and Horsehead) as well as for NGC2023, the contour shapes reveal a similar L-shaped morphology in the $G_0-P_{\text{th}}$ parameter space, contrasting with the more extended contour of the intensely irradiated Orion Bar. 
While the minima are relatively well defined (only a narrow region in parameter  space yields fits comparable to the best model), these L-shapes indicate some non-trivial degeneracy between $G_0$ and $P_{\text{th}}$ constrained by H$_2$ ro-vibrational lines.

Figure \ref{fig:Excitation} \rev{displays} the excitation diagrams for S140, IC63, the Horsehead nebula, NGC2023, and the Orion Bar. The \rev{symbols show} IGRINS observations with their associated uncertainties, \rev{alongside colored semi-transparent lines indicating a factor of 2 margin}. 
The black lines for S140 and IC63 and the grey dashed lines for NGC2023, the Horsehead and the Orion Bar correspond to the best-fit model obtained without additional constraints (red dot in Fig. \ref{fig:contour}).
\rev{Finally, solid} black lines for the Horsehead, NGC2023 and the Orion Bar correspond to the best-fit model \rev{obtained after inclusion of additional constraints as discussed in the next section}.

The five PDRs are well-reproduced by the Meudon PDR code, with most modeled line intensities falling within a factor of two of the observed values. This quantitative agreement, assessed by the reduced $\chi^2$ for the unconstrained models indicates that the Meudon PDR code accurately captures the relevant physical and microphysical processes in PDRs detailed in Section \ref{Method}, with particular emphasis on the excitation and de-excitation processes that populate H$_2$'s rovibrational levels. A detailed investigation of the populating processes for every ro-vibrational level of H$_2$ is discussed in detail in Section \ref{Section:OPR}.

Despite this overall agreement, we observe a systematic overestimation of para-H$_2$ level populations (triangles) at high rotational quantum numbers ($J$) for all PDRs. This is particularly evident for $v=1$ (in green in Fig. \ref{fig:Excitation}) at high $J$ ($J>8$) in IC63, S140 and NGC2023. This suggests that the Meudon PDR code may not fully account for a process influencing the intensity of these para lines and consequently the ortho-to-para ratio in the gas and is discussed in Section \ref{Section:OPR}.

For S140, an excellent overall fit is achieved, with the exception of the aforementioned overestimation of a few high-J para-H$_2$ lines. We add that the geometrical scaling factor $\Omega$ of the upper values of the most confident model is at the acceptable limit of 10. Therefore, a closer model might exist for higher values of $\Omega$  but we deem a geometrical scaling factor exceeding 10 unrealistic, given that the best model with a relaxed scaling factor remains similar to the initially proposed best model.

Concerning IC63, the best-fit model accurately reproduces the observations, except for a few high-J para-H$_2$ lines. The geometrical scaling factor $\Omega$ is 3.5, which is reasonable given the complex geometry of IC63. 

For the three other PDRs, further investigations were made thanks to additional constraints from previous studies, as discussed in the next section. The excitation diagrams are therefore going to be discussed in the next section by comparing the unconstrained best-fit model (grey dashed line) and the constrained best-fit model (black solid line) in Figure \ref{fig:Excitation}.

\subsubsection{Constrained fitting results}

To reduce the aforementioned degeneracy for the low-excitation PDRs and detail the large region in parameter space for the Orion Bar, we consider some additional strong constraints based on previous studies of the well-studied PDRs Orion Bar, Horsehead and NGC2023. 

Table \ref{Table:BestFit} summarizes the 5 most relevant models proposed in this article, with the same two unconstrained models form IC63 and S140, and with the three constrained models for the Horsehead, NGC2023 and Orion Bar PDRs. The 68\% confidence intervals (CI) on the three parameters are also provided, derived from the posterior distribution obtained with \textsc{Beetroots}. The $\chi^2_R$ values and the percentage of lines reproduced within a factor of 2 are also given for these best-fit models.

\begin{figure*}[htbp!] 
    \centering
    \vspace{-0.5em}     \includegraphics[width=\linewidth]{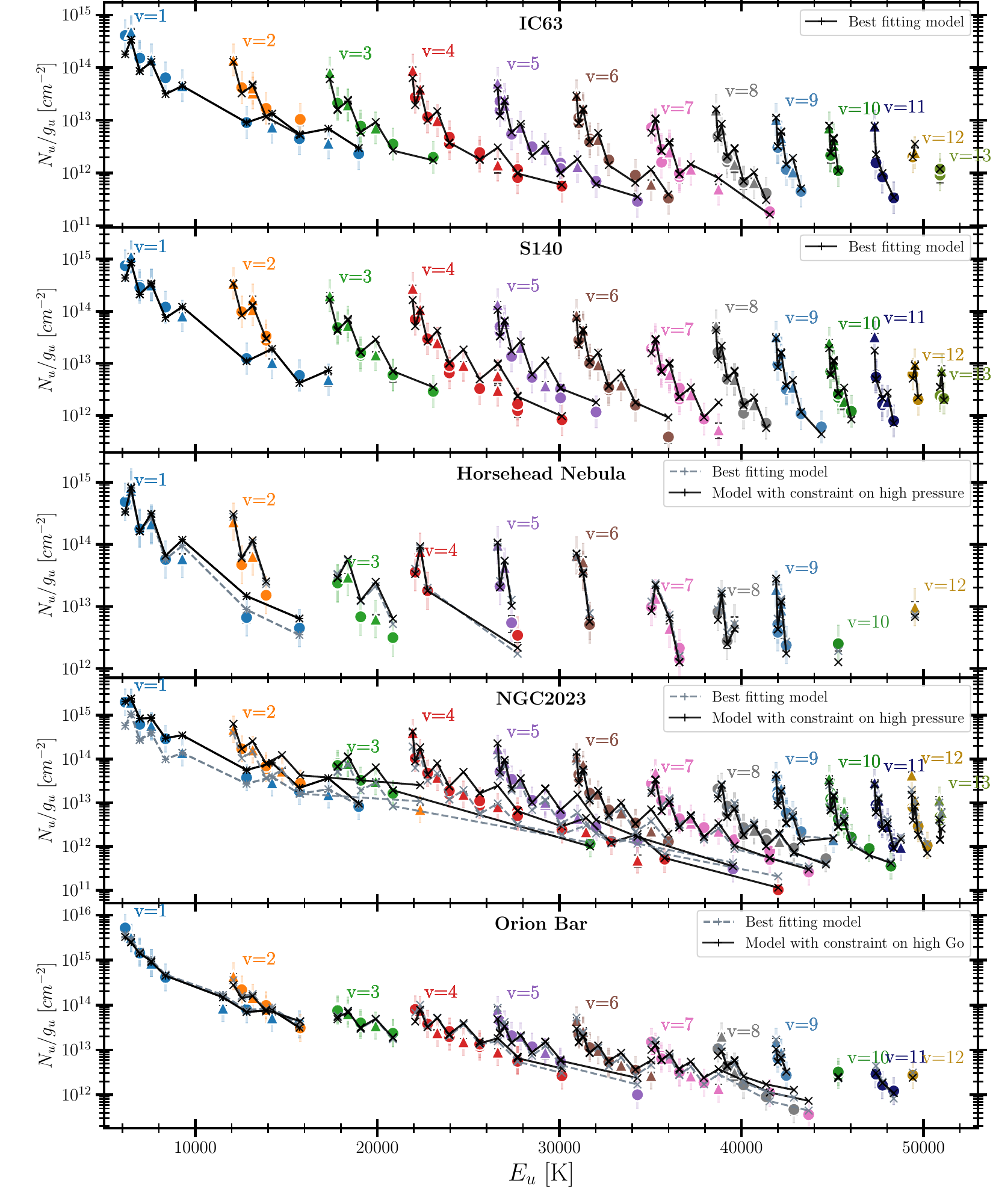}
         \caption{Excitation Diagram of (a): IC63, (b): S140, (c): Horsehead Nebula, (d): NGC2023, (e): Orion Bar with IGRINS observations in dot (ortho levels) and triangles (para levels) with observational uncertainties from modeling uncertainties and in semi-transparent the error bars within a factor of 2. In \rev{black} is the best fitting model without uncertainties and in \rev{grey} the best fitting model under the constraint for the Horsehead Nebula, NGC2023 and Orion Bar. The plain lines are the most relevant models presented in Table \ref{Table:BestFit}.}
        \label{fig:Excitation}
\end{figure*}

Concerning the Horsehead Nebula, recent high-resolution observations from ALMA have provided constraints on the spatial scale of the atomic region \citet{Hernandez2023} found that the distance between the ionisation front and the H/H$_2$ transition could not be resolved at the resolution of the observations resulting in an upper limit of 650 AU. 
The black line in Fig.~\ref{fig:contour} panel (c) represents models with the H/H$_2$ transition at exactly 650 AU. The shaded region at low-pressure corresponds to models exceeding this limit and therefore not suitable. The green cross represents the minimum $\chi^2_R$ model satisfying this constraint. \rev{As seen in the third panel of Fig.~\ref{fig:Excitation}, this constrained best-fit model (black continuous line) performs almost as well as the unconstrained one (grey dotted line). The pressure increase required to satisfy the spatial constraint primarily results in a poorer fit for high-$J$ levels of $v=1$. Additionally, both models seem to have similar difficulties reproducing the $v=3$ levels}.

The parameters of the best-fit model for NGC2023 appear to be in disagreement with those deduced by \citet{Sheffer2011} from Spitzer H$_2$ observations. Given the L-shape degeneracy of the parameters for NGC2023, we explored whether different models giving good fits to our IGRINS observations could also reproduce the Spitzer observations of \citet{Sheffer2011}. We show in Appendix~\ref{NGC2023} that a model at the tip of high-pressure branch in our L-shaped degeneracy branch also provides a good fit to the Spitzer observations (while the other tested models do not). We thus deduce from this additional constraint that the high-pressure branch of the L-shaped degeneracy is to be favored, and propose approximate values $P_{\text{th}} = 5 \times 10^7\,\text{K}\,\text{cm}^{-3}$ and $G_0 = 10^3$ as more likely to be representative of the conditions in NGC2023.
\rev{As shown in the third panel of Fig.~\ref{fig:Excitation}, this high-pressure model performs almost as well as the unconstrained fit,} \rev{successfully reproducing over 85\% of the lines within a factor of 2 with a $\chi^2_R = 3.6$}. \rev{This high pressure model behaves very similarly to the best fit model on high $v$ lines ($v\ge6$), but seems to favor matching the ortho levels on lower $v$ states, while the best fit model favored matching the para levels. Both appear unable to match the ratio between successive ortho and para levels except for the lowest $J$ states in each $v$ level.}

The broader contours for the Orion Bar PDR indicate less sensitivity of H$_2$ to the physical parameters in this high-excitation PDR. The numerous previous studies of this PDR have quite firmly established that $G_0>10^4$ \citep{Tielens1985,Marconi1998,Peeters2024}. In order to better constrain the parameters, we thus restrict our search to the domain satisfying this constraint. 
The green cross on Fig.~\ref{fig:contour} marks the resulting constrained best-fit model derived from the Bayesian estimation with strong prior on constraint and exclusion of models according to the constraints.
\rev{As shown in the fifth panel of Fig.~\ref{fig:Excitation}, this constrained model ($\chi^2_R = 2.26$) fits the observations reasonably well compared to the unconstrained one ($\chi^2_R = 1.25$), with only a slight degradation for high-$J$ levels at $v > 4$.}

Overall, the hundreds of \rev{observed} H$_2$ rovibrational lines \rev{cannot} simultaneously constrain $G_0$ and $P_{\text{th}}$ \rev{on their own. This limitation manifests as} narrow, high-pressure and high-$G_0$ solution branches \rev{for the Horsehead and NGC2023}, and as a flatter optimum \rev{for the Orion Bar}. \rev{However, incorporating prior observational knowledge, such as constraints on the atomic region size, or on the FUV flux effectively breaks these degeneracies in the $G_0-P_{\text{th}}$ plane. For all three cases, the constrained models remain accurate, showing only a slight degradation in $\chi^2_R$ and in the percentage of lines reproduced within a factor of 2 compared to the unconstrained best-fits. Incorporating external physical priors alongside H$_2$ observations thus leads to more robust and precise parameter estimates without compromising the overall fit quality.}

\rev{The detailed spatial profiles of the modeled PDRs, for quantities such as species densities and gas temperature, are presented in Appendix \ref{Appendix:Profile}.}

\subsection{Excitation of H$_2$ at formation on dust surfaces}\label{Section:Sizun}

A long-standing question regarding H$_2$ is to determine the specific levels in which it forms and whether it is possible to detect a signature of this excitation in observations (see \citet{Burton2002} for a tentative detection). The formation of a molecule of H$_2$ releases 4.5 eV, which is partitioned among internal energy, translational energy and the grain itself, to overcome the binding energy and to heat the grain. \citet{Duley1986} suggested that a part of the energy goes into vibration of the nascent molecule with a peak at $v = 6-7$, an energy close to the difference between the binding energy and the depth of the potential well on amorphous silicate surfaces. \citet{Black1987} tested three energy partitioning scenarios: the \citet{Duley1986} prescription, a case where all formation occurs in a single vibrational level ($v = 14$) with energy corresponding to the total available formation energy, and a simpler scenario assuming equipartition among rotational, vibrational, and translational energies. They demonstrated that these different scenarios lead to variations in the predicted H$_2$ infrared line intensities, implying that observations of the 1–2 $\mu$m lines could provide valuable information on the excitation state of nascent H$_2$ and on grain properties. 
Indeed, different formation scenarios predict distinct signatures in the rovibrational level populations of H$_2$. A model considering H$_2$ formation in $v=6-7$ would enhance those observed levels and reduce the intensity of high rotational levels of $v=0-1$. We therefore tend to expect differences at low $v$, high $J$ and maybe at high $v$ as well.

Experimental studies conducted on cold surfaces, which predominantly probe mechanisms related to physisorption, found that H$_2$ is preferentially formed in vibrationally excited levels, around $v$ = 4, \citep{Islam2010, Lemaire2010}, with a lower fraction of the formation energy going to rotational excitation. At the edges of PDRs, where grains are warm, the formation of H$_2$ via chemisorbed H atoms may play a significant role, notably through the Eley–Rideal mechanism. Using classical molecular dynamics calculations combined with accurate zero-point energy (ZPE) motions for chemisorbed H on graphene, \citet{Sizun2010} found that H$_2$ formed via the Eley–Rideal mechanism also exhibits a higher fraction of the formation energy going to vibrational modes, with a peak around $v = 5-6$ and only moderate rotational excitation. Very recently, \citet{Jubert2025} used \textit{ab initio} molecular dynamics simulations to determine the vibrational state of formation of H$_2$ on graphene slabs via different formation mechanisms. Also they find that H$_2$ is formed in high vibrational states.

In our models, H$_2$ is formed at the edges of PDRs via the Eley–Rideal mechanism, and we therefore focus on this process. The public version of the Meudon PDR code only accounts for the repartition of the formation energy between internal, kinetic and grain derived by \citet{Sizun2010}, but assumes a Boltzmann distribution for the internal energy, with a mean energy corresponding to  the \citet{Sizun2010} internal energy fraction.
While \citet{Sizun2010} computed the complete distribution of nascent H$_2$ across the $v, J$ levels, their paper only presents the marginal distributions of $v$ and $J$ graphically, and the tabulated data corresponding to these results is unavailable.
For the present work, we extracted an approximate $v, J$ distribution directly from Fig. 1 of \citet{Sizun2010}, which loses any information about the correlations between $v$ and $J$ and present the two distribution in Figure \ref{fig:SizunDistrib} in Appendix \ref{App:Sizun}.
In the present section, we compare the models extracted from \citet{Sizun2010} to models using the default prescription (Boltzmann distribution) of the public version of the Meudon PDR code.

\begin{figure*}[h!]
    \centering
        \begin{tabular}{cc}
    \includegraphics[width=0.48\textwidth]{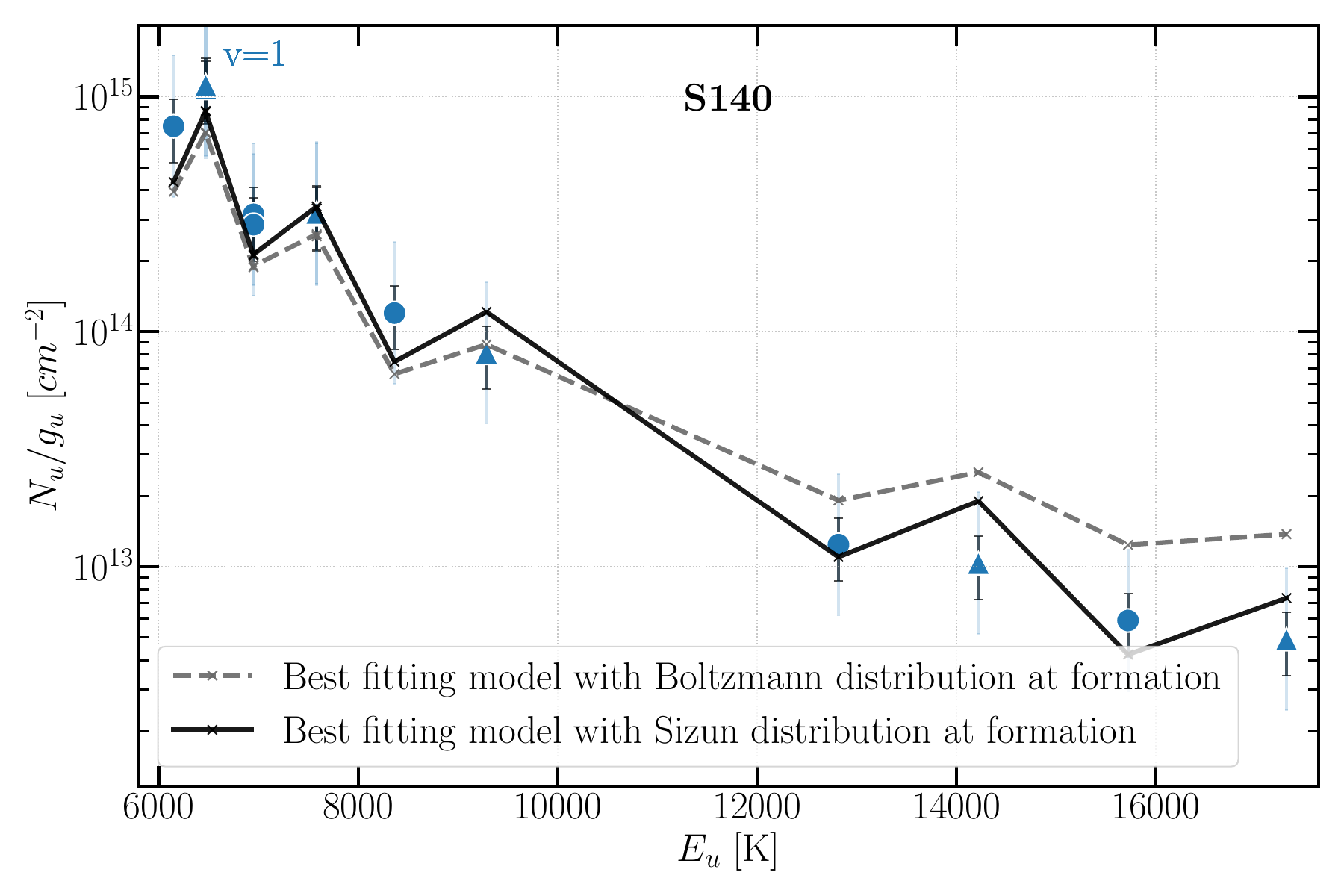}  &  \includegraphics[width=0.48\textwidth]{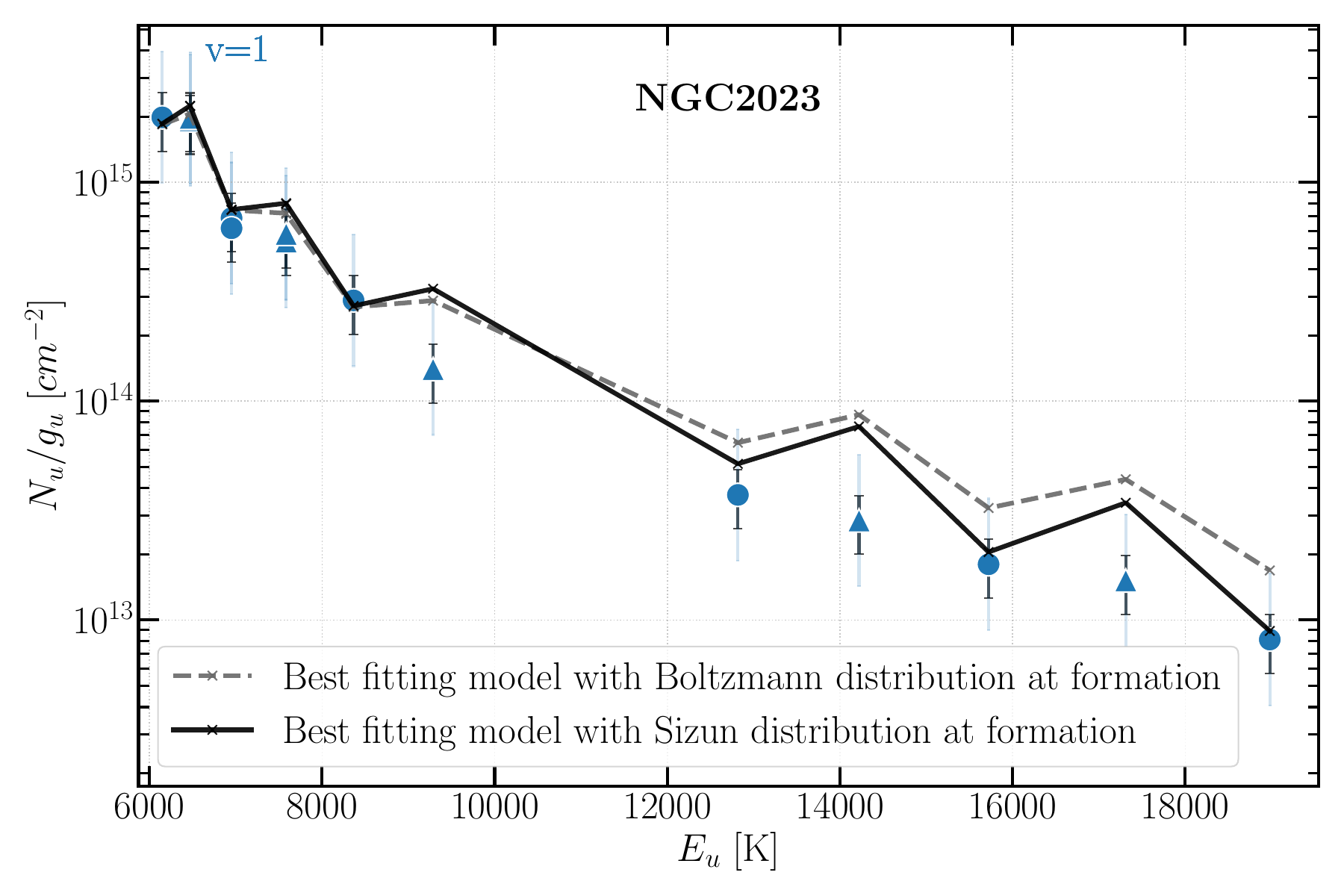}
        \end{tabular}
      \caption{H$_2$ excitation diagram for $v$ = 1 for S140 and NGC2023. \rev{IGRINS observations are represented} \rev{as discrete symbols}. \rev{The triangles represent the para lines and the dots the ortho lines}. The line called "Boltzmann distribution" corresponds to a simplification \citet{Sizun2010} where nascent H$_2$ is formed following a Boltzmann distribution at a temperature corresponding to 2.7 eV (see text) and the line labelled "Sizun distribution" corresponds to the new $v, J$ distribution we extracted from \citet{Sizun2010}. The 3 other PDRs are in Appendix \ref{App:Sizun}.}
  \label{fig:Sizun}
 \end{figure*}

\rev{Figure~\ref{fig:Sizun} illustrates how H$_2$ excitation at formation impacts the modeled line intensities. We highlight S140 and NGC2023 as they represent the two extremes of our low- to intermediate-excitation regimes: S140 features the lowest $G_0$ and pressure, whereas NGC2023 acts as an intermediate case with high pressure and $G_0 \sim 10^3$. Analogous results for the other PDRs are provided in Appendix~\ref{App:Sizun}.}

For the low excitation PDRs such as the Horsehead, S140, IC63 and NGC2023, we find that models with the default Boltzmann prescription tend to overestimate the high-$J$ levels in particular in $v=1$, while including the \citet{Sizun2010} level distribution significantly reduces the discrepancy by improving the overall $\chi^2_R$ of $v=1$ by 44\% for the Horsehead, 19\% for IC63, 43\% for S140 and 30\% for NGC2023.

This shows that the observation of these emission lines contain information about the excitation of nascent H$_2$ on grains by the Eley-Rideal mechanism. Additional tests performed using different $J$ distributions show that distributions in which a significant fraction of nascent H$_2$ is formed in high rotational levels are not compatible with the observations as they tend to result in overpopulated high-J levels in $v=1$. Indeed, while H$_2$ is mostly formed in higher $v$ levels, the ro-vibrational cascade modifies only slowly the $J$ distribution due to the $\Delta J= -2;0;+2$ selection rule for radiative transitions.
The IGRINS observations studied in this article thus suggest that H$_2$ with most of its internal energy in vibration, and are thus consistent with the theoretical predictions of \citet{Sizun2010}.
No significant difference is observed for higher vibrational levels, which are mostly populated by FUV pumping and subsequent fluorescence.

In the case of the dense and strongly illuminated PDR of the Orion Bar (see Fig. \ref{fig:SizunOthers} in Appendix \ref{App:Sizun}), we find no significant difference in the fit quality between models using the two prescriptions.This might be caused by excitation and de-excitation processes occurring more rapidly and thus erasing the memory of the initial formation levels.
\rev{Analysis of JWST observations toward the Orion Bar PDR provides further evidence for low rotational and high vibrational excitation of freshly formed $\text{H}_2$ on dust grains \citeauthor{MeshakaInprep} (in prep.).}

\section{Discussion}	\label{Discussion}
\subsection{H$_2$ ro-vibrational excitation processes in PDRs}\label{Processes}

The Meudon PDR code reproduces the H$_2$ line intensities in its ro-vibrational levels quite satisfactorily. This is thanks to to a detailed treatment of the physics of H$_2$, including the calculation of non-LTE excitation in the levels of the ground electronic state, coupled with chemistry that incorporates a detailed description of formation on grains and photodissociation. Indeed, this process is directly treated at the level of the FUV radiative transfer algorithm that enables a precise determination of FUV pumping rates level by level  taking into account a rigorous treatment of line shielding and then fluorescence or dissociation.

\begin{figure*}[htbp!]    
    \subfloat[Horsehead Nebula]{\includegraphics[width=0.5\textwidth]{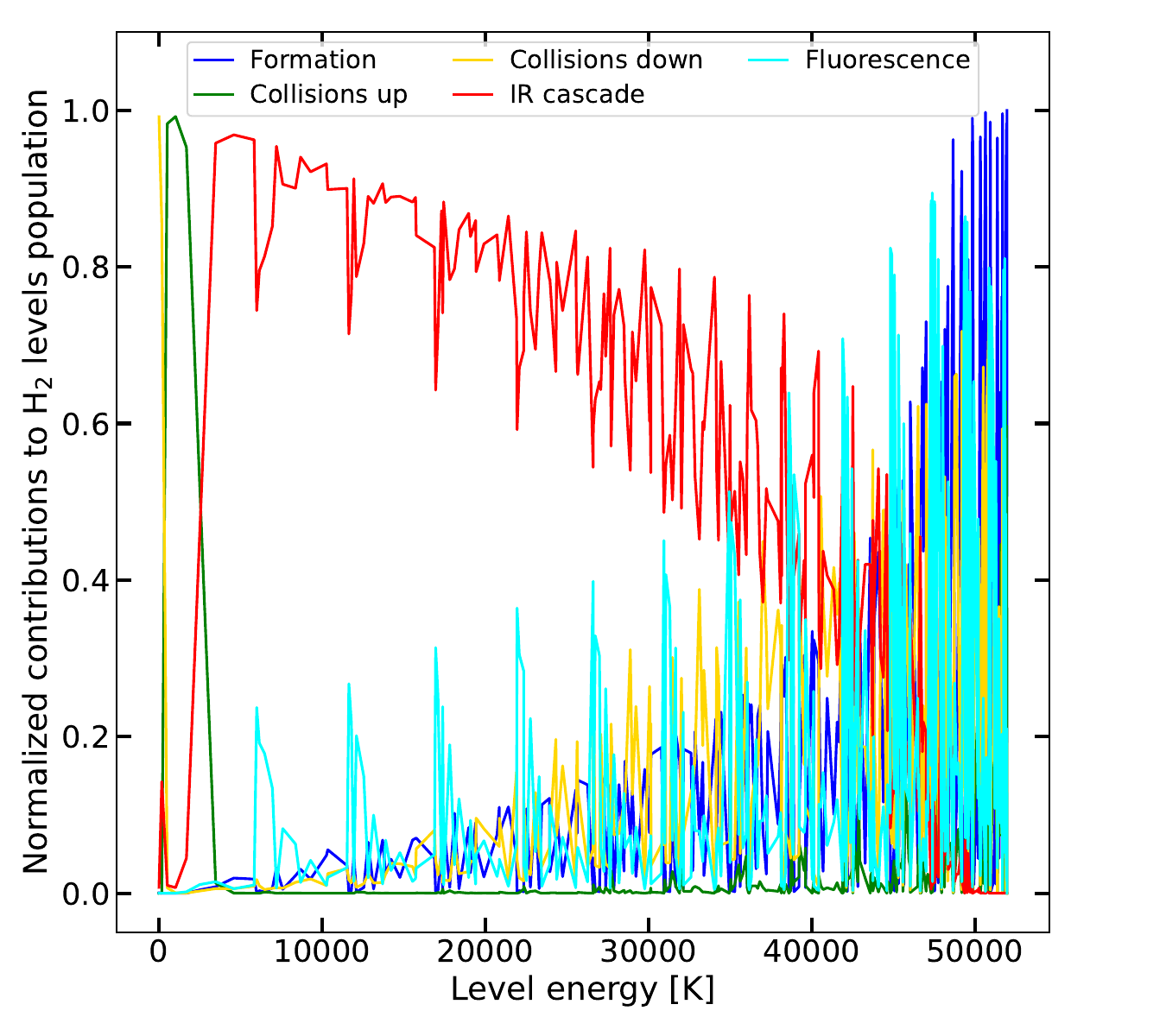}}
    \hfill
    \subfloat[Orion Bar]{\includegraphics[width=0.5\textwidth]{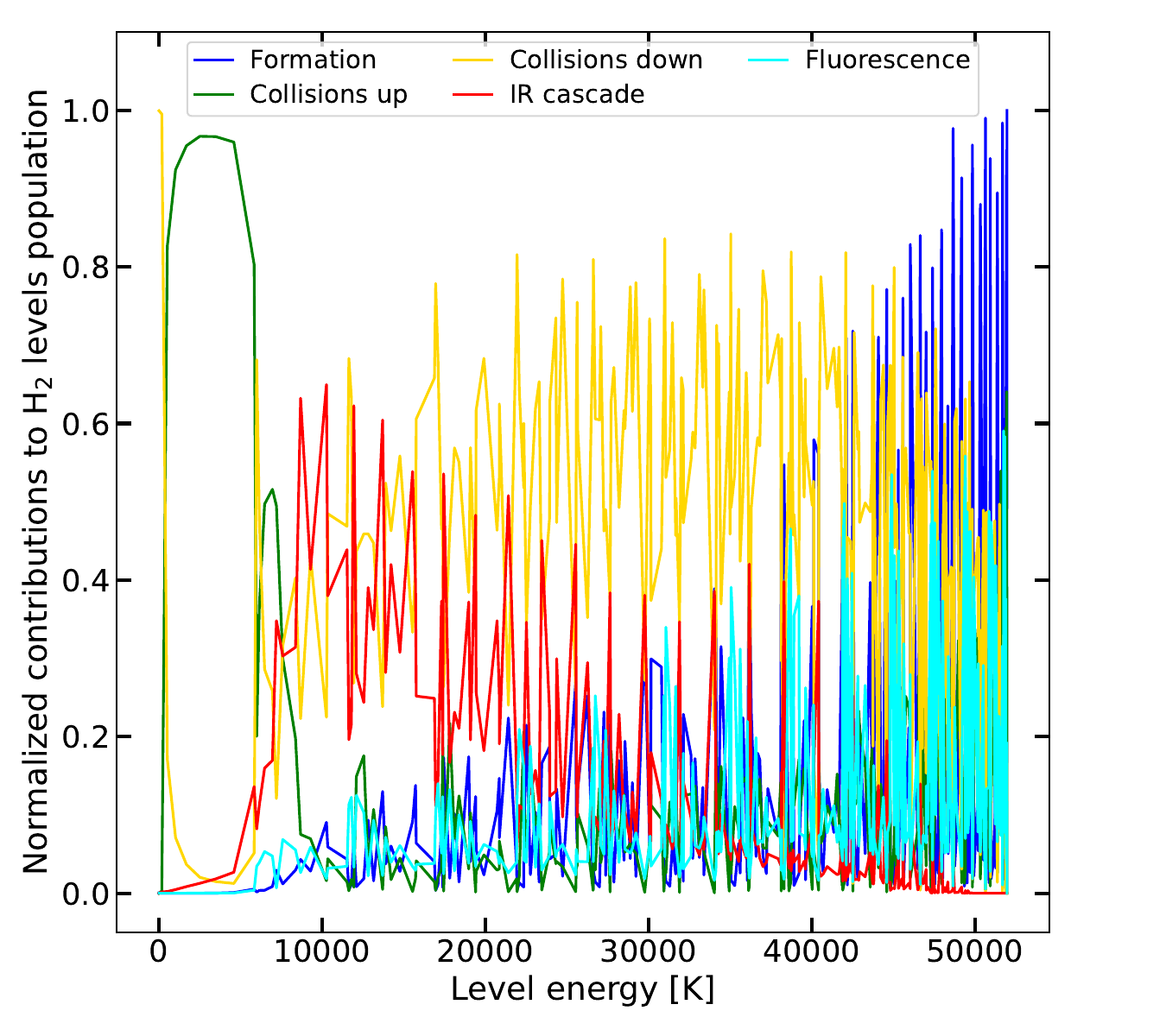}}
    \caption{Comparisons of the processes populating the H$_2$ levels for  the Horsehead Nebula  and the Orion Bar best models from Table~\ref{Table:BestFit}. The rates of the processes are normalized for each level. On the x axis, the levels are organized by increasing energy. These processes are evaluated at the H/H$_2$ transition position, where the emissivity of ro-vibrational lines reaches its maximum ($A_{\mathrm{V}} \simeq 1.25$ mag for the Orion bar, and $A_{\mathrm{V}} \simeq 0.5$ mag for the Horsehead Nebula).}
    \label{fig:H2_Domin_process}
\end{figure*}

We analyzed the processes responsible for the excitation of the various ro-vibrational levels of H$_2$ in two extreme cases, the Orion Bar and the Horsehead nebula. Figures~\ref{fig:H2_Domin_process} presents the relative contributions of the processes responsible for populating each H$_2$ level. These processes include the direct excitation at formation, upward and downward collisional transitions, and spontaneous radiative de-excitation, referred to hereafter as the IR cascade (with IR absorption and stimulated emission being negligible compared to the other mechanisms), as well as FUV-pumping followed by fast radiative decay, referred to as fluorescence. In these figures, the relative populating rates for each level are displayed as a function of the level energy.
        
As expected from their low energies and the radiative selection rules, the populations of the first levels are controlled by collisions. Conversely, the highest energy levels of the ground electronic state are predominantly populated by injection from electronically excited states via fluorescence and by direct excitation at formation. In the intermediate regime, a more complex behavior emerges. In the Horsehead Nebula, these intermediate levels are mostly populated through radiative IR cascade: that is, they are fed by de-excitations from higher energy states that were initially populated ("pumped") either by fluorescence or by formation processes, which then decay stepwise along the ro-vibrational ladder. In contrast, in the Orion Bar where the density is larger than in the Horsehead, this cascade is driven predominantly by collisional transitions rather than by radiative ones. This is an important point to note. It is often assumed that collision rates are significant only for low-lying levels, which is why they are seldom calculated for high-energy states. Only recently have physicists computed collisional rates for high-$v$, high-$J$ levels of H$_2$ \citep{Bossion2018, Wan2018}. As demonstrated here, access to such rates is essential for interpreting H$_2$ observations in these high $v$, $J$ levels. It is only for levels below 12,000 K ($v$ = 2, $J$ = 12) that IR cascade dominates collisions. 

We thus emphasize that in dense PDRs such as the Orion Bar, de-excitation is largely dominated by a collisional cascade for a wide range of high-energy levels, while lower energy levels are governed by radiative transitions. In contrast, in less dense environments like the Horsehead Nebula, the ro-vibrational cascade remains radiative across a much broader energy range.

\subsection{Comparison with previous studies}\label{sect:comparison_to_litterature}
	
In this section, we discuss the parameter estimates we have obtained and compare them with previous studies that solely relied on H$_2$ lines for parameter constraints. \rev{We note that as the H$_2$ emission used as constraint in this study arises from a narrow layer close the H/H$_2$ transition (cf. Appendix~\ref{Appendix:Profile}), the thermal pressures we estimate are strictly speaking only representative of this layer. If a pressure gradient were to exist in the PDR, pressure estimates derived from tracers of other layers of the PDR might be expected to differ.}
The distances between the illuminating star and the PDR front deduced from the minimisation procedure are also provided in Table~\ref{Table:BestFit}.
These values can be compared with the lower limits provided by the projected distances listed in Table~\ref{Table:PDRRef} (the corresponding $G_0$ values are also represented in Fig.~\ref{fig:contour} as the red dotted lines). 
The distances we obtain are fully compatible with these lower limits for Orion Bar and NGC2023.
However, models for low excitation PDRs (IC63 and Horsehead) tend to require distances that are too short by less than a factor of 2.
This suggests that the model requires more \rev{FUV} illumination than physically available to accurately reproduce the lines. We find that improved modeling of the incoming radiation field geometry and spectral shape (as detailed in Appendix~\ref{Spectrum}) significantly reduces the needed $G_0$ to model these PDRs, although not sufficiently to satisfy the projected distance constraints.

In S140 and IC63, the initial pressure estimates were derived solely from the rough density and temperature estimations presented in Table~\ref{Table:PDRRef}. While these early estimates provided a good approximation of the resulting pressure, our new models offer a significant improvement, tightening the uncertainty to a factor of 1.5, compared to the previous order of magnitude uncertainty.

Concerning the Horsehead Nebula, the resulting pressure of $P = 5-7 \times 10^6\, \text{K} \,\text{cm}^{-1}$, which is slightly superior to the pressures proposed in \citet{Hernandez2023} and equivalent to the one derived in \citet{Zannese2025} that were based on more approximate density and temperature estimations. This higher pressure originates from the additional constraint implemented concerning the distance from the ionisation front to the H/H$_2$ transition, therefore taking into account dynamical effects of compressions as detailed in the following section. We note that very recent JWST observations \citep{Abergel2024, Zannese2025} seem to further reduce the upper limit on the distance between the ionization front and the H/H$_2$ transition to $\sim100$ AU. We kept the upper limit at $650\ \text{AU}$ \citep{Hernandez2023} in order to avoid an overly constrained prior in our model and to be able to explore the parameter range. 
A tighter constraint would tend to push the best-fit model toward even higher pressures.

In NGC2023, \citet{Sheffer2011} concluded to a high $G_0 \sim 10^4$) and high density ($n_\text{H} \sim  10^5$ cm$^{-3}$, which combined with the observed H$_2$ excitation temperature results in $P_\mathrm{th} \sim 3.5 \times 10^7$ K cm$^{-3}$) to fit pure rotational H$_2$ observations, but did not quantify the parameter range capable of accurately reproducing the observations. With the current version of the Meudon PDR code and our more realistic \rev{FUV} illumination setting, a lower $G_0 \approx 10^3$ and slightly higher pressure ($P_{\text{th}}\approx 10^8 \, \text{K}\, \text{c}m^{-3}$) 
is able to reproduce the pure rotational H$_2$ observations of \citet{Sheffer2011} while simultaneously providing a good fit of the ro-vibrational lines observed with IGRINS (cf. Appendix~\ref{NGC2023}). While the high $G_0$ value initially proposed by \citet{Sheffer2011} is incompatible with more recent estimates of the distance of NGC2023 from Earth \citep{GaiaCollaboration2021}, our result is consistent with the resulting projected distance constraint.

In the Orion Bar, our model favors a $G_0$ value that is consistent with, but at the lower end of, previously estimated values \citep{Marconi1998, Peeters2024}, and a pressure lower by a factor of 2 to 5 compared literature values \citep{Joblin2018,Goicoechea2016}.
The fact that the pressure estimated from H$_2$ ro-vibrational lines is lower than values reported in CO studies as in \citet{Joblin2018}, could be understood as a consequence of the fact that CO and H$_2$ do not originate from the same layers within the PDR, with rotational CO lines emitted in deeper layers. Moreover, recent observations of HD in the Orion Bar \citep{Zannese2025} also find lower pressure.
The discrepancy suggests some physical variation, unaccounted for by our model, between the emission layers of the two molecules.
Possible variations could be a pressure gradient associated with dynamical effects, or variations in dust properties across the PDR.
It should however be kept in mind that increasing the pressure and radiative intensity would likely have only a moderate impact on the current best $G_0$-constrained model, as Fig.~\ref{fig:contour} shows that the $\chi^2_R$ surface is relatively flat in that region.

\subsection{Correlations between physical parameters}

\rev{A correlation between the intensity of the incident FUV field and the thermal pressure \citep{,Joblin2018,Wu2018,Seo2019,Cormier2019,Palud2025}  \revA{or similarly between the UV field intensity and density \citep{YoungOwl2002}} of various galactic and extragalactic PDRs has been observed using different tracers , as discussed in the review by \citet{Wolfire2022}. The relationship would follow a near  linear relationship with $G_0\propto P_{th}$ in \cite{Joblin2018}, $G_0\propto P_{th}^{0.9}$ in \cite{Wu2018} and $G_0\propto P_{th}^{0.75}$ in \cite{Seo2019}. Two distinct phenomena could account for the existence of such a relationship between the two main parameters impacting the PDR modeling. }

\rev{The first proposition would be the photo-evaporation phenomenon driven by radiative feedback compressing the edges of the molecular cloud}. Far-UV photons emitted by massive stars heat the gas at the ionization and dissociation fronts, driving to an expansion of the local region of the cloud, that in turn exerts additional pressure on the surrounding neutral gas \citep{Bertoldi1996, Bron2018}. This relatively high-pressure region, illuminated by FUV photons, becomes the site of distinctive photochemistry. The FUV radiation supplies the energy needed to overcome activation barriers, enabling the formation of molecules near the cloud edges \citep{Goicoechea2016, Goicoechea2017, Joblin2018, Goicoechea2019}. Also, the photo-evaporation can induce dynamical effects that can shift the H/H$_2$ transition closer to the edge of the PDR \citep{Maillard2021}. 

\rev{A second explanation, proposed in \citet{YoungOwl2002} and  \citet{Seo2019}, would be that if we assume pressure equilibrium between the PDR and a static H$\textsc{ii}$ region, Str\"omgren scaling laws result in a a pressure proportional to $G_0^{3/4}$. This illustrates that a $P_\mathrm{th}-G_0$ relationship in PDRs might not necessarily imply the presence of photoevaporation dynamics, although the applicability of Str\"omgren scalings in the blister-type H$\textsc{ii}$ regions that most archetypical PDRs border remains questionable.}

\begin{figure}
  \includegraphics[width=0.5\textwidth]{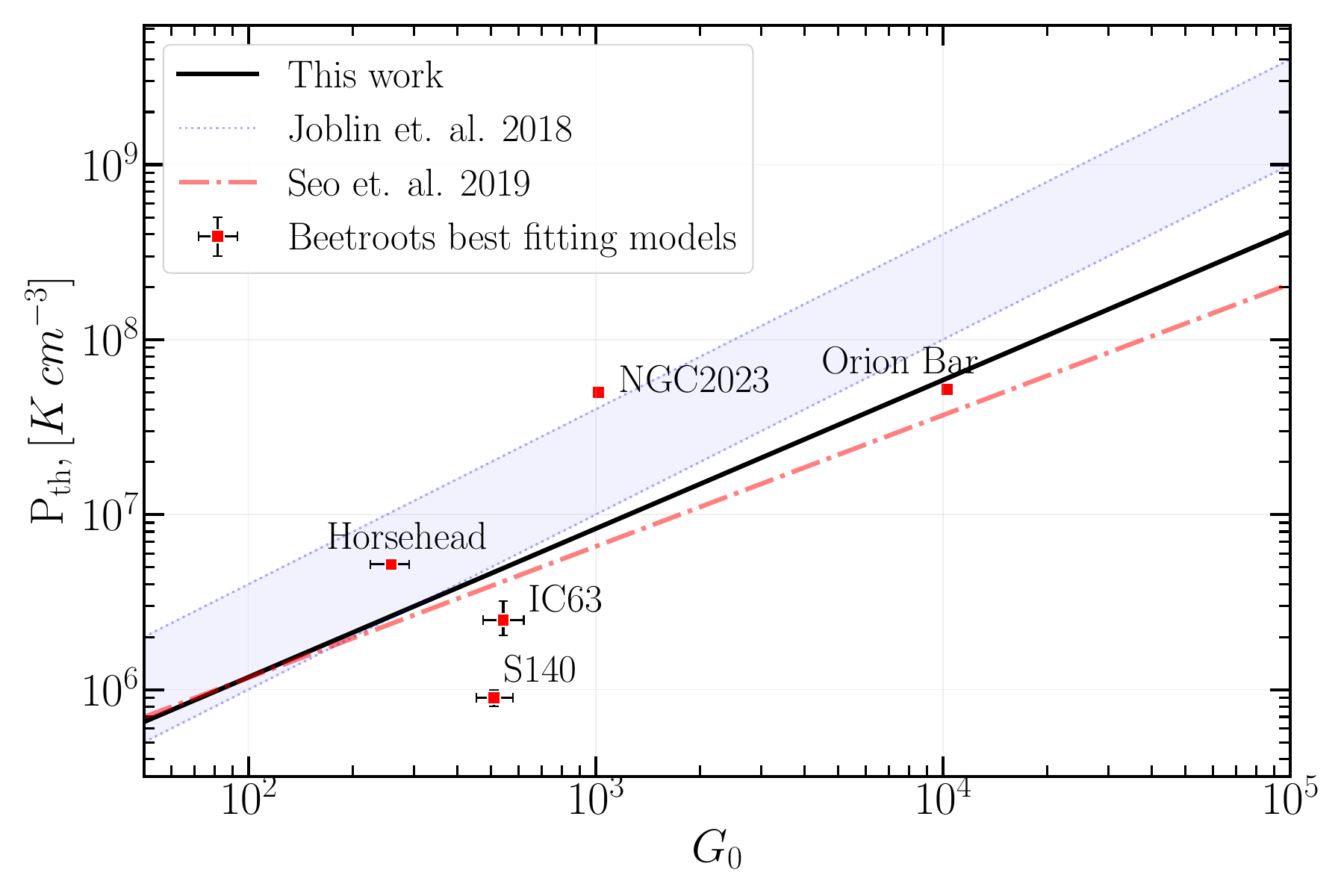}
  \caption{Relation between the thermal pressure in the PDRs studied here and the FUV intensity $G_0$. \\
  The dashed blue lines show the range of values obtained by \citet{Bron2018} in their photo-evaporating models. 
  The orange semi dashed lines represent the relationship obtained in \citet{Seo2019} 
  The black line represents this works power law regression as detailed in the text.} 
  \label{Fig:P_Go}
\end{figure}

Figure~\ref{Fig:P_Go} shows a scatter plot of our $P_\mathrm{th}$ and $G_0$ estimations for the five PDRs studied in this paper, compared to several previously proposed relationships. Although our sample is small and shows a significant scatter in the $P_\mathrm{th}$-$G_0$ plane, our result seems consistent with these power-law relationships. Based on our results, we find a best power-law fit $P_{\text{th} } =2.3\times10^4\times G_0^{0.85}$ that is consistent with the aforementioned relations.

\rev{Our fit lies between the theoretical behaviors predicted by photo-evaporation models and by pressure equilibrium with the adjacent ionized region. Because of the low number of PDRs studied in the present work and the significant scatter of the estimates in the $P_{\text{th}}$--$G_0$ plane,  the current data does not clearly favor one explanation over the other. Studying a larger sample of PDRs would be necessary.}

The addition of observational constraints seems to favor a power-law relationship with an exponent less than one, rather than a linear relationship, although the small sample size and the scatter on the results make it difficult to conclude.
This flatter slope seems to be in part a consequence of obtaining lower pressures in the high-excitation PDR of the Orion Bar using H$_2$ ro-vibrational lines, than from other previous constraints such as high-J CO lines, as discussed in Sect.~\ref{sect:comparison_to_litterature}.
The scatter in the pressure values obtained in low excitation PDRs may be due to the absence of additional constraints applied to IC63 and S140, as we found that in the case of the Horsehead PDR, the addition of spatial scale contraints favored the high-pressure solution within the degeneracy in the fit from H$_2$ line intensities alone.


\subsection{Ortho-Para ratio}\label{Section:OPR}

\begin{figure*}[ht!]
  \centering
  \includegraphics[width=0.95\textwidth]{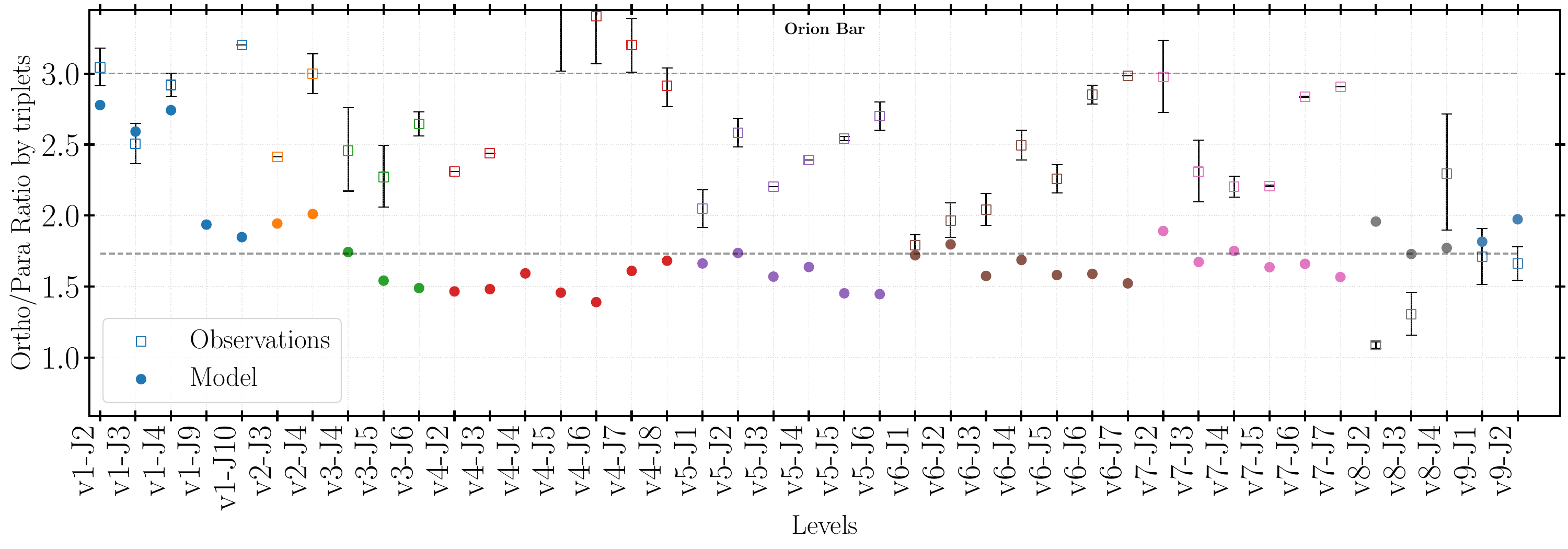}
  \caption{Ortho-to-para ratio calculated on 3 consecutive levels from the IGRINS observations of the Orion Bar following the procedure described in \citet{Bron2016}. The x-axis displays the middle level from the set of three used to derive the ortho-para ratio. Grey dashed line are for $\sqrt{3}$ and 3.} 
  \label{fig:OPR}
\end{figure*}

As stated in Section~\ref{Results}, the great amount of detected lines allows for a detailed study of micro physical processes such as those controlling the ortho-to-para ratio (OPR) of $H_2$. 
While the OPR of H$_2$ in the lower $J$ levels of $v=0$ and its deviation from LTE has been studied before based on ISO and Spitzer observations \citep{Fuente1999,Moutou1999,Habart2003,Habart2011,Fleming2010,Bron2016}, IGRINS observations give an unprecedented view on the ortho-para processes affecting higher vibrational and rotational levels.
We calculated the OPR for each triplet of H$_2$ lines (3 lines coming from consecutive upper $J$ at same upper $v$) in both the model and the observations, following the methodology described in \citet{Bron2016}. This method assesses the deviation of the intermediate level column density from the slope defined by the first and third levels of each triplet.

The figures representing the OPR by triplet for the four low excitation PDRs are shown in Appendix~\ref{OPR} whereas the most interesting case of the Orion Bar is presented in Fig.~\ref{fig:OPR}.
Empty squares represent the OPR derived from the observational data, while filled circles indicate the OPR from the most relevant model (green cross in Fig.~\ref{fig:contour}), with consistent color-coding across Fig.s~\ref{fig:contour} and~\ref{fig:Excitation}, and separated by vibrational level. 
For the observations, the error bars only represents the scatter on the level column density deduced from multiple lines arising from the same emitting level, when several such lines are detected.
It should be noted that there are additional sources of uncertainty not represented here, which is why some OPR values are found to exceed 3—something that is not physically possible, at least within the context of these observations of H$_2$ emission lines in PDRs.

For the low excitation PDRs (S140, Horsehead and IC63 in Fig.~\ref{fig:OPRIntermediate}, the derived mean OPR of the models fits the observational OPR relatively well, although we observe a systematic increase of the OPR in high-J levels (in particular for v=1) that is not explained by our models.
For the other levels, the OPR is mostly between 1 and $\sqrt{3}$.
Indeed, \citet{Sternberg1999} showed that in the vibrational levels, an OPR equal to $\sqrt{OPR_\text{gas}}$ (i.e. the OPR including all vibrational levels, largely dominated by the lower $J$'s in $v=0$ is 
expected if the populations of the vibrational levels are dominated by UV pumping followed by fluorescence and purely radiative cascade. 
In low excitation PDRs, previous ISO and Spitzer observations have shown that the OPR in $v=0$ is significantly lower than 3 and often close to 1, so that the mechanism of \citet{Sternberg1999} predicts an OPR in the vibrational level between between 1 and $\sqrt{3}$ as found both in the IGRINS observations and in the models.

On the other hand in the Orion Bar, the OPR in the vibrational levels of H$_2$ deduced from the observations is significantly higher than 1 and often close to 3.
Due to the higher density conditions in the Orion Bar, the lower $J$ levels of $v=1$ are populated by collisional excitation (cf. Fig.~\ref{fig:H2_Domin_process}), and their OPR is close to the thermal value of 3 as a result, both in the observations and in the models. For higher ro-vibrational levels however, significant discrepancies appear between the model and the observations: while the model predicts OPR values close to $\sqrt{3}$, OPR values derived from the observations are mostly higher than $\sqrt{3}$, reaching values close to 3 for the highest observed $J$s of most vibrational levels.

As the true OPR of the gas in the Orion Bar is close to 3, the value of $\sqrt{3}$ predicted in the models is expected if collisional ortho-para conversion is inefficient during the cascade that follows UV pumping and fluorescence. Otherwise, conversion would tend to push the OPR towards 3 during the cascade. However, despite the inclusion of recent collisional data for high-$v$ levels in the Meudon PDR code, our model results show that collisional deexcitation during the cascade (which dominate the cascade for the higher $v$'s as shown on Fig.~\ref{fig:H2_Domin_process}) do not provide sufficiently fast ortho-para conversion to significantly change the OPR from its value of $\sqrt{3}$ (caused by optical depth effect of the pumping lines). The discrepancy with the observations thus remains unexplained. We have also explored the influence of other processes: ortho-para conversion on dust surfaces and the OPR of nascent H$_2$ molecules both seem to have negligible impact on the OPR of the observed vibrational levels in the conditions of the Orion Bar PDR.

\subsection{\rev{Uncertainties on FUV extinction}}\label{extinction_discussion}

\rev{As mentioned, in Sect.~\ref{model_description} and \ref{incident_UV_fields}, simplifying assumptions concerning extinction in our models were made: 1) we neglected extinction within the H$\textsc{ii}$ region when estimating the FUV field incident on the PDR, 2) we assumed a fixed extinction curve within the PDR material, both across all studied PDRs and across depth within a single PDR.}

\rev{Dust extinction within H$\textsc{ii}$ regions has been hard to quantify due to contamination by neutral gas components directly associated with the H$\textsc{ii}$ region (neutral shell, embedded dense clumps, ...) or lying in the foreground. It is theoretically expected that radiation pressure on dust grains inside H$\textsc{ii}$ region both redistributes the material towards the external regions creating an inner cavity \citep{Draine2011}, and causes dust drift that can expel a significant fraction of the dust into the surrounding neutral shell \citep{Akimkin2017,Ishiki2018} reducing the dust-to-gas mass ratio inside the H$\textsc{ii}$ region by a factor 2 to 10. Such a scenario of partial dust depletion due to radiative pressure was also advocated by \citet{Paladini2012} based on an observational study of dust IR emission in a sample of evolved H$\textsc{ii}$ regions. In addition, the smallest grains (PAHs or nano-grains) might be depleted due to photodestruction \citep{Pilleri2012,Pilleri2015}. Such a PAH depletion in the ionized gas has been observed in the LMC \citep{Paradis2011} and more recently across 42 nearby star-forming galaxies \citep{Egorov2023,Egorov2025}. Destruction of the smallest grain would result in a very flat extinction curve in the visible-UV range, further reducing the total FUV extinction contribution from the H$\textsc{ii}$ region. Modeling of the observed dust emission close to the ionization front in the Orion Bar indeed indicates a depletion of nano-grains by a factor 15 and  an extinction curve completely flat in the visible and UV range \citep{Elyajouri2024}\footnote{\rev{A similar study by \citet{Elyajouri2025} in the Horsehead nebula found no significant nano-grain depletion in the Horsehead, but contrary to the Orion Bar, the high compactness of the Horsehead PDR makes it harder to separate dust emission coming from the ionization front and from denser PDR material.}}.}
\rev{As a result, we have neglected the impact of dust extinction in the ionized gas when building the incident FUV field. If significant, such extinction would result in a reduced $G_0$ reaching the ionization front. As the extinction curve seems to be very flat in H\textsc{ii} regions, the hardness $H^\mathrm{LW}$ of the FUV field would be less affected.}

\rev{Similar processes, with the addition of collision-driven fragmentation of larger grains, are likely to make the dust population in the PDR layer, and thus the local extinction properties, dependent both on the environment and on the depth within the PDR \citep{Schirmer2022}. Variations of extinction properties across depth in the PDR can have significant impacts on the PDR structure and observables \citep{Goicoechea2007}. However, the lack of a quantitative theory relating the dust population to the local conditions within PDR layers makes it difficult to include such a variation in the model. Observational determinations of extinction properties within PDR material are also scarce. For ease of comparability, we chose to use a single extinction curve for all the PDRs studied here, using the same prescription as \citet{Joblin2018}.
According to the theory of the H/H$_2$ transition of \citealt{Sternberg2014}, dust FUV extinction controls the H/H$_2$ transition only  under conditions where $\alpha G > 1$  corresponding roughly to $G_0/n_\mathrm{H} > 10^{-2}$ cm$^{-3}$. H$_2$ emission in the low-excitation PDRs of our sample would thus be unaffected by changes in the extinction properties, as they have negligible dust extinction at the H/H$_2$ transition. More significant effects could be expected in high-excitation PDRs such as the Orion Bar.}

\section{Conclusion}

In this work, we exploit \rev{existing} IGRINS observations of tens of H$_2$ ro-vibrational lines from $v=1$ to $v=13$ toward five PDRs (the Horsehead Nebula, IC63, S140, NGC2023, and the Orion Bar) to constrain the physical conditions \rev{($G_0$ and thermal pressure)} and processes \rev{(distribution of nascent H$_2$ on grains)} within these regions. To this end, we employed the Meudon PDR code to model each source, searching for the best \rev{FUV} radiation field strength $G_0$ and gas thermal pressure $P_{\text{th}}$, that explain the observations. We computed grids of models using irradiation by stellar spectra tailored to each specific PDR and applied two complementary statistical approaches to assess the agreement between the models and the observations: $\chi^2$ minimization and Bayesian inversion (using Beetroots, \citealt{Palud2025}).

First, our PDR models succeed to reproduce most of the observed H$_2$ line intensities. This agreement is made possible thanks to a detailed modeling of the physics of H$_2$ in the Meudon PDR code, the implementation of recent atomic and molecular data, and a detailed modeling of the spectral shape and geometry of stellar illumination on the PDR. Particular emphasis is placed on the necessity of considering collisional transitions for H$_2$ even for high ro-vibrational levels, especially in strongly irradiated PDRs. Indeed, we found that in dense PDRs, collisional de-excitations are the dominant processes populating the ro-vibrational levels of H$_2$ in the range $E_u\approx 15 000 - 45 000$ K .

The H$_2$ ro-vibrational lines alone are found to be insufficient to fully constrain the physical conditions ($G_0$ and P${_\text{th}}$) as degeneracies persist. However, incorporating additional constraints intrinsic to each PDR can help breaks these degeneracies, enabling a more precise determination of the \rev{FUV} flux intensity and the gas thermal pressure.
The resulting physical parameters and their corresponding uncertainties for these constrained models are listed in Table~\ref{Table:BestFit}.

\rev{In addition, we found that careful modeling of the incident FUV field  was necessary to model properly each of the PDRs. Using a scaled ISRF spectrum, as often done in PDR modeling, or an isotropic irradiation geometry, is found to bias the $G_0$ estimate towards higher values, resulting in apparent irreconcilable discrepancies with other constraints such as the projected star-PDR distance.
 Using a normal illumination geometry, combined with either a specific stellar spectrum, or an equivalent blackbody spectrum with an effective temperature consistent with the illuminating star,
results in a near identical fit of the H$_2$ line intensities but in lower $G_0$ estimates, closer to the geometrical distance constraints.
We note that the resulting incident FUV spectra have significantly different shapes compared to the ISRF spectral shape often used in PDR modeling. We thus propose to characterize this difference by the FUV hardness parameter $H^\mathrm{LW}$ representing the ratio of the energy density within the $912\,\text{\AA}$ - $1108\,\text{\AA}$ range to that in the $912\,\text{\AA}$ - $2400\,\text{\AA}$ range.}

This systematic study of five typical PDRs allows us to investigate the P$_\text{th} - G_0$ relationship using H$_2$ as tracer for the first time. We find a quasi-linear correlation consistent with the CO-based results of \citet{Joblin2018}. We recall that such a relationship \rev{can be} expected when photo-evaporation induced by the FUV radiation field compresses the edges of molecular clouds \citep{Bron2018} \rev{but can also highlight a pressure equilibrium with the adjacent H$\textsc{ii}$ region \citep{Seo2019}.}  
Extending this analysis to a larger sample of PDRs observed in H$_2$ is necessary to further refine this relationship \rev{and determine which phenomenon this relationship illustrates.}

For the first time, these data allow to constrain, observationally,  the ro-vibrational excitation of nascent H$_2$ at formation on grain surfaces. In PDRs with moderate physical conditions (Horsehead Nebula, S140, IC63 and NGC2023), we find that observed populations of high-J levels of the $v=1$ vibrational state require an H$_2$ formation scenario in which molecules are formed with high vibrational but low rotational excitation as predicted theoretically by \citet{Sizun2010}. This signature is not observed in dense PDRs such as the Orion Bar, where the nascent rovibrational distribution is likely erased by collisional transitions.

We also analyzed the ortho-to-para ratio in each $v$ level of H$_2$ observed in the 5 PDRs. For low excitation PDRs, we find the ortho-to-para ratio in vibrationally excited levels to be mostly consistent with the pure radiative fluorescence scenario described by \citet{Sternberg1999}, with values between $\sqrt{3}$ and 1 (due to the true OPR, dominated by $v=0$ being lower than 3).
In the case of the Orion Bar, we find that the observations show high values of the OPR (up to 3), not only in $v=1$ (where collisional excitation leads to a thermalized OPR), but also in higher $v$ levels, contradicting our model results that fall back to $\sqrt{3}$ as predicted by a radiative fluorescence scenario in this case. Despite using the most recent collisional deexcitation rates for H$_2$ in vibrational levels, collisions appear unable, in our models, to thermalize the OPR in the cascade that follows FUV pumping and fluorescence.

Our results highlight the extensive information that can be obtained by studying H$_2$ ro-vibrational lines over a wide range of vibrational levels, which is now possible with instruments such as NIRSpec on the JWST and the ground-based NIR spectrograph IGRINS.

\begin{acknowledgements}

This work used The Immersion Grating Infrared Spectrometer (IGRINS) which was developed under a collaboration between the University of Texas at Austin and the Korea Astronomy and Space Science Institute (KASI) with the financial support of the US National Science Foundation under grants AST-1229522, AST-1702267 and AST-1908892, McDonald Observatory of the University of Texas at Austin, the Korean GMT Project of KASI, the Mt. Cuba Astronomical Foundation and Gemini Observatory.\\ This work was supported by the Thematic Action “Physique et Chimie du Milieu Interstellaire” (PCMI) of INSU Programme National “Astro”, with contributions from CNRS Physique \& CNRS Chimie, CEA, and CNES. This research was achieved using the POLLUX database (pollux.oreme.org) operated at LUPM (Université de Montpellier - CNRS, France) with the support of the PNPS and INSU. The authors are grateful to the anonymous referee for their constructive remarks.    
\end{acknowledgements}

\addcontentsline{toc}{section}{References}
\bibliographystyle{aa} 
\bibliography{Piluso2026_H2}

\appendix

\section{Presentation of the 5 studied PDRs}\label{Appendix:PDRs}

\rev{This appendix details the specific properties of the five PDRs in our sample, providing the necessary context for our modeling without cluttering the main body of the paper.}

The S140 PDR lies at the interface between the L1202/L1204 molecular cloud and the S140 H\textsc{ii} region that surrounds the B 0.5V star HD 211880.
This star radiates with $T_{\mathrm{eff}}=29,000$ K at a projected distance of approximatively 1.85 pc, or 7', \citep{Poelman2005} from the PDR surface located at the north east of the star, resulting in a rather weak $G_0$ of approximately $100$ \citep{Timmermann1996}. The S140 PDR is seen nearly edge-on.
Previous modeling studies based on CO and $^{13}$CO observations \citep{Spaans1997}, on C91$\alpha$ and C$^{18}$O observations \citep{Wyrowski1997} and on ISO H$_2$ rotational lines observations \citep{Timmermann1996,Habart2004} have resulted in an estimated gas density of $1-5\times 10^4$ cm$^{-3}$ and an estimated incident FUV intensity $G_0 \sim 100 - 400$. Combining these density estimates with the rotational excitation temperature of H$_2$ of 400K \citep{Habart2004} yields a thermal pressure near the H/H$_2$ transition of $P_\mathrm{th}= 2 - 10 \times 10^6$ K cm$^{-3}$.
The IGRINS observations detected 131 rovibrational lines of H$_2$ in the S140 PDR.

IC63 is a reflection nebula mainly illuminated by the star $\gamma$ Cas, located at a projected distance of 1.3 pc from the PDR \citep{Andrews2018}. $\gamma$ Cas has been studied in \citet{Sigut2007} and appears to be a B0.5IV star with $T_{\textrm{eff}}=25,000$ K with some uncertainty on the stellar mass (13 to 18 $M_\odot$). \citet{Andrews2018} deduced an incident $G_0$ on the PDR of $\sim 150$ from dust SED observations with Herschel and Spitzer, although a more recent analysis by \citet{Eiermann2024} combining UV-optical and IR observations suggested an incident $G_0$ of only $38-58$.  
The most recent PDR studies based on ISO \citep{Habart2004,Thi2009} and Spitzer \citep{Fleming2010,Andrews2018} observation of rotational H$_2$ emission have found a range of gas densities $n_\mathrm{H}\sim10^3 - 10^4$~cm$^{-3}$ and a H$_2$ rotational temperature of $\sim 700$ K in the warm component, from which we deduce a thermal pressure $P_\mathrm{th} \sim 3.5\times10^5 - 3.5\times10^6$~K~cm$^{-3}$.
This source has also been studied in H$_2$ UV fluorescence lines by \citet{France2005}. The IGRINS observations detected 120 H$_2$ rovibrational lines in the IC63 PDR.

The Horsehead PDR is one of the most famous and well studied PDR and is often used as the archetype of low-excitation PDRs (PDRs with $G_0$ of at most a few hundreds).
The Horsehead Nebula is a portion of the Orion B molecular cloud and appears as a dark column against the bright emission of the IC 434 H\textsc{ii} region in visible wavelengths. 
The main star illuminating the PDR is the O 9.5V star $\sigma$ Ori Aa, which is the brightest component of a triple hierarchical system with $\sigma$ Ori Ab and $\sigma$ Ori B \citep{SimonDiaz2015}. 
This system is estimated to be at a projected distance of d $\approx3.5$ pc from the PDR, resulting in $G_0 \sim 100$ \citep{Abergel2003}.
The Horsehead PDR has long been studied from the radio, revealing in particular a rich chemistry \citep{Guzman2012, Pety2012, Guzman2013}, to the near-infrared domain. H$_2$ observations with Spitzer \citep{Habart2011} lead to an estimated density of $n_\mathrm{H}=10^4$~cm$^{-3}$, with a rotational temperature varying from 264 to 747 K across the detected $J$ levels, which corresponds to a thermal pressure $P_\mathrm{th} \sim 1.3 - 3.2 \times 10^6$~K~cm$^{-3}$. Based on ground-based high spatial resolution observations of 1-0 S(1) rovibrational line of H$_2$ with SOFI, \citet{Habart2005} concluded that the PDR density profile was best described as close to isobaric with $P_\mathrm{th}\sim 4\times 10^6$~K~cm$^{-3}$. More recently, ALMA observations constrained the extremely thin size of the atomic region, while resolving the separation between the H/H$_2$ and the C$^+$/C/CO transition, and comparisons with PDR models constrained $P_\mathrm{th} \sim 3.7-9.2 \times 10^6$~K~cm$^{-3}$.
Recent observations with the JWST \citep{Abergel2024} have unveiled a highly detailed structure of the PDR, highlighting fine sub-structures such as 600 au large filaments and striations perpendicular to the illumination front. 
The IGRINS observations detected 53 H$_2$ rovibrational lines in the Horsehead PDR.

NGC2023 is a reflection nebula located in Orion B molecular cloud and surrounding the B1.5V star HD~37903. This $T_{\textrm{eff}}=23,000$ K star carved out a small spherical cavity within the L1630 molecular cloud \citep{Compiegne2008}. HD 37903's stellar parameters are not well constrained, we thus used the prescription of \citet{Kurliliene1981} to derive approximate value of stellar radius and mass from its spectral type. 
Our IGRINS observations focus on the bright southern ridge.
Previous estimations of the physical conditions in the southern ridge of NGC2023 based on H$_2$ rotational emission observed with ISO \citep{Habart2004} and Spitzer \citep{Fleming2010,Sheffer2011} resulted is $n_\mathrm{H}\sim 10^4 - 10^5$~cm$^{-3}$ with a rotational temperature of H$_2$ of $\sim690$~K, corresponding to a thermal pressure $P_\mathrm{th}\sim 3.5 - 35 \times 10^6 $~K~cm$^{-3}$. The $G_0$ irradiating the PDR was estimated in \citet{Sheffer2011} to be $\sim 10^4$ using the projected distance, and assuming a distance from Earth to HD~37903 of $350\pm50$~pc. A more recent distance for Gaia EDR3 \citep{GaiaCollaboration2021} is $399\pm4$~pc. Combined with the possibility that the true star-PDR distance is greater than the projected distance, this makes $G_0$ of a few thousands also plausible.
The IGRINS observations detected an impressive number of 175 rovibrational lines, reaching $v$=13 in NGC2023.

The Orion Bar is also one of the most famous and well studied PDRs and is usually used as the archetype for high-excitation PDRs (PDRs with $G_0 \gtrsim 10^4$). 
The Orion Bar is a dense ridge of molecular gas and dust forming a prominent diagonal structure at the southeastern edge of the Orion Nebula H  {\small II} region surrounding the Trapezium star cluster, of which the main contributor is $\theta^1$ Ori C \citep{Goicoechea2019}.
This star is the hottest and the heaviest among the stars illuminating the PDRs considered in the present study, with $T_{\textrm{eff}}=39,000$K and M $>$ 30 $M_\odot$ according to \citet{SimonDiaz2006}.
The Orion Bar has long been a testing ground for PDR models. Previous studies of the Orion Bar with Herschel \citep{Joblin2018} and ALMA \citep{Goicoechea2016,Goicoechea2017} concluded to thermal pressures of $1-3 \times 10^8$~K~cm$^{-3}$. A range of FUV irradiation with $G_0 = 1-3\times 10^4$ was proposed by \citep{Marconi1998} and has been used in numerous studies since.
Recent JWST observations of the Orion Bar have significantly advanced our understanding of its structure and physical processes. 
The observations revealed a complex, 3D "terraced" geometry and fine substructures, filaments and ridges at both large and small scales \citep{Habart2024, Peeters2024}. 
The NIRSpec observations reveal a rich spectrum characterized by recombination lines (He {\small I}, H {\small I}, Fe {\small II}, ...), fluorescence lines (O {\small I}, N {\small I}, ...), aromatic infrared bands, and rovibrational emission lines of H$_2$, HD, and CO, based on which \citet{Peeters2024} have proposed a range of $G_0$ values of $2.2-7.1\times10^4$.
Recent NIRSpec data have allowed the detection of 78 high-$J$ H$_2$ lines (up to $J=17$ for $v=0$ and $J=19$ for $v=1$), extending up to the $v=6$ vibrational level \citep{Peeters2024}, whose PDR analysis is underway \citep{MeshakaInprep}. 
Meanwhile, IGRINS observed in the Orion Bar 101 H$_2$ rovibrational lines reaching $J=12$ and $v=13$, providing a more in-depth exploration of the higher vibrational levels of H$_2$.

\section{Incident FUV field in models}\label{Spectrum}

PDR models often model the irradiation of the PDR  by the Interstellar Standard Radiation Field (ISRF, \citealt{Mathis1977,Draine1978}) scaled by a factor to represent the presence of nearby O and B stars. However, the spectrum of a massive star differ substantially from the ISRF spectral shape. 
Furthermore, models sometimes keep the isotropic nature of the ISRF while modeling PDRs illuminated by one dominant star or cluster. In such case, the  incident radiation field on the PDR should be nearly unidirectional. 
In this Appendix, we study the impact of geometrical orientation and spectral shape of the incoming radiation field.

\subsection{Illumination orientation}
\rev{We compared two model grids in which the FUV illumination is represented by scaled-up ISRFs: one assuming isotropic illumination and the other assuming a beamed radiation field incident normal to the PDR surface.}
We find the model predictions for the H$_2$ lines to be shifted to lower $G_0$ values when switching from isotropic to normal illumination, so that, for example, the best-fit $G_0$ for the Horsehead PDR decreases from 880 (isotropic) to 450 (normal) for a nearly identical match to the observations. This factor of $\sim 2$ arises because FUV photons penetrate deeper into the cloud under normal incidence. \rev{This illustrates that, in the literature, the inferred values of $G_0$ depend on the geometry of the FUV radiation field adopted in the models.}

\subsection{Illumination spectral shape}

 \begin{figure}
  \centering
  \includegraphics[width=0.48\textwidth]{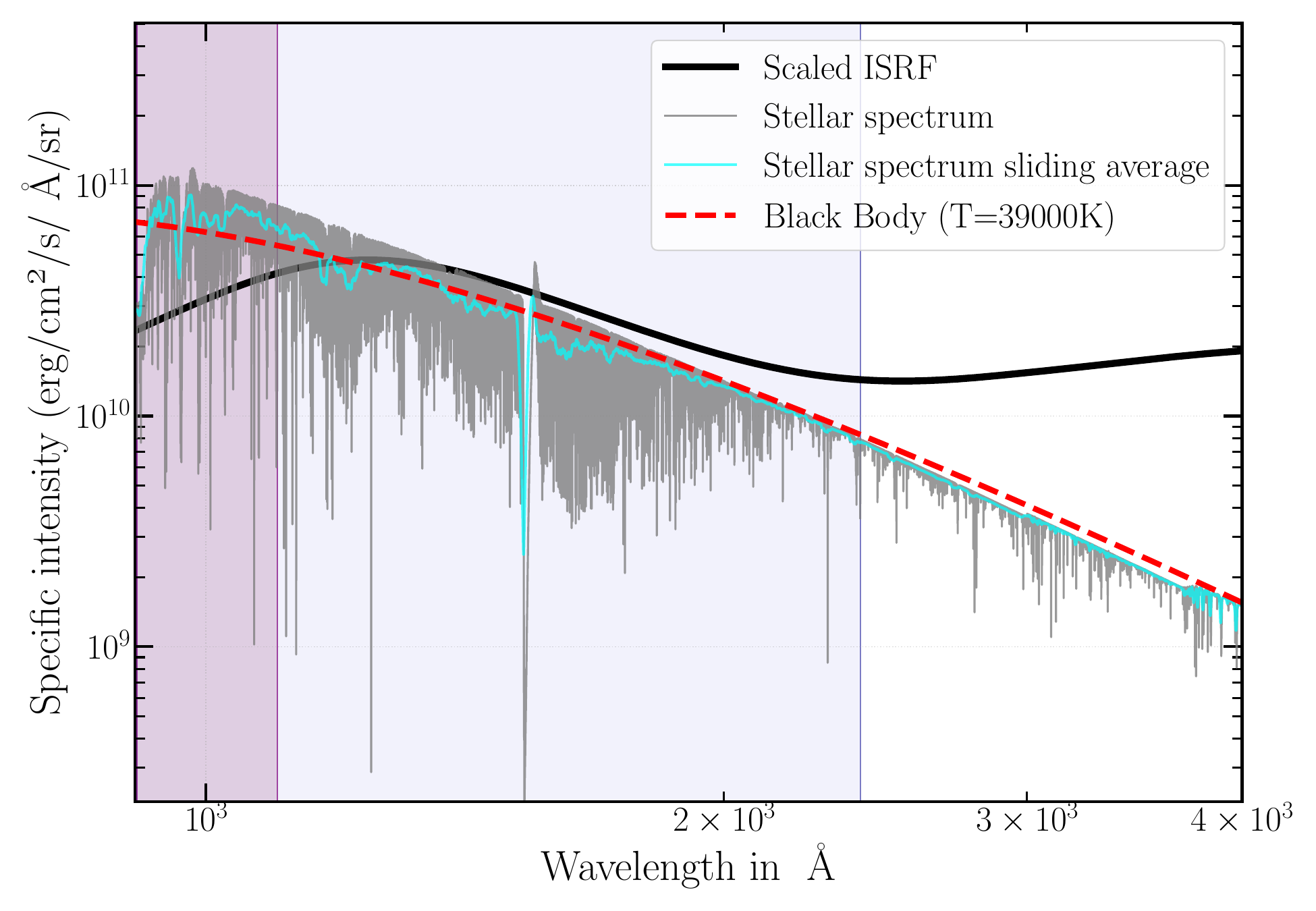}
  \caption{Specific intensity of the different illumination spectra used in the study at same $G_0$ for the Horsehead Nebula. In black: Mathis ISRF scaled to fit the $G_0$, in grey the theoretical spectrum used to represent $\sigma$ Ori Aa, in blue its sliding average and in red the equivalent blackbody spectrum at $T=T_{eff}+6000$ K. In deep purple is the 912--1100~\AA\ section and in light purple the 912--2400~\AA\ section of the spectrum. }
  \label{fig:Spectra}
\end{figure}

Figure~\ref{fig:Spectra} compares the Mathis ISRF to the stellar spectrum chosen for the Horsehead PDR ($T_{\text{eff}} = 33,000~\text{K}$, see Table~\ref{Table:Stars}) from the Pollux Database \citep{Palacios2010}. Both spectra are normalized to yield the same $G_0$ over the 912--2400~\AA\ range at the PDR surface. 
The shape of the Mathis ISRF differs significantly from the stellar continuum in the 912--1100~\AA\ range that is the Lyman-Werner band where H$_2$ electronic excitation and photo-dissociation take place. In this band, the ISRF intensity is significantly lower than the stellar spectrum. \rev{The hardness parameter $H^\mathrm{LW}$ defined in Eq.~\ref{eq:Hardness} is 0.15 for the Mathis ISRF, while it is 0.316 for the stellar spectrum considered here.}
To quantify the impact on modeling results, we compared two identical model grids, both with normal illumination but changing only the spectral shape from that of the Mathis ISRF to that of the stellar model selected for the Horsehead PDR (cf. Table~\ref{Table:Stars}).
Our study reveals that using the ISRF spectral shape requires a $G_0$ value twice as high to produce similar H$_2$ predictions as a model using the stellar spectrum, as a result of the underestimated intensity in the 912--1100~\AA\ range. \rev{At first order, the H$_2$ emission predictions thus seem to depend on the product $H^\mathrm{LW}\, G_0$ rather than on $G_0$.}

Additional tests also have also shown that neglecting stellar emission and absorption lines has no impact on the final modeling results, as long as the total intensities in the relevant spectral ranges are equivalent: using smoothed version of the stellar spectrum (using a sliding average) yields nearly identical model predictions for the H$_2$ lines, and using a blackbody spectrum also yields equivalent results as long as its temperature is adjusted to have an equivalent fraction in the 912--1100~\AA\ range. The presence or absence of stellar lines in the incident spectrum is not critical, at least to first order.

As a consequence, for the sake of accuracy, the analysis presented in this article only used models illuminated with a normal geometry and a specific stellar spectrum selected for each individual PDR (cf. Table~\ref{Table:Stars}).


\section{NGC2023 case}\label{NGC2023}

\begin{figure}[htbp] 
  \centering
  \includegraphics[width=0.5\textwidth]{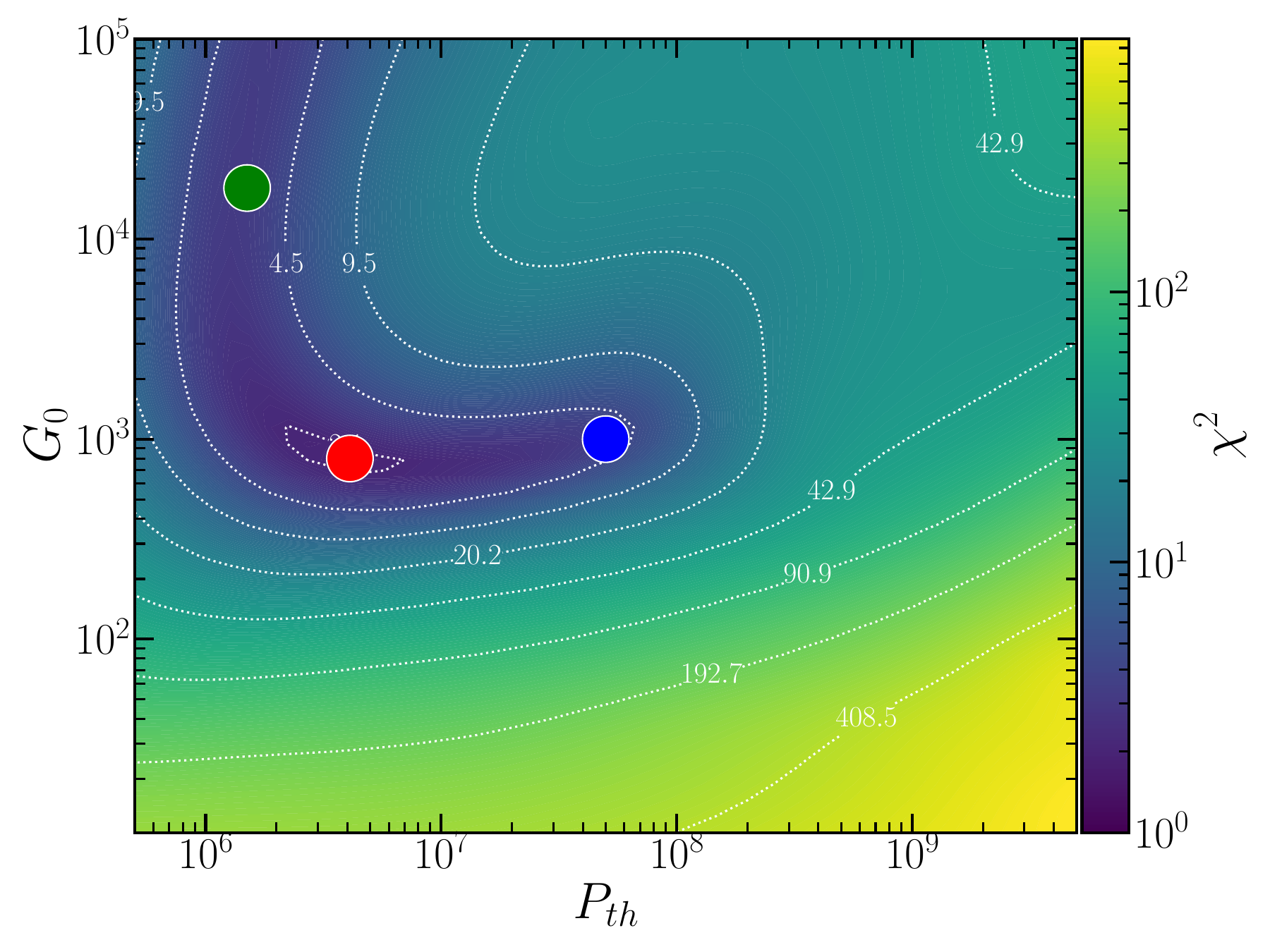}
  \caption{$\chi^2$ contour map of NGC 2023 model grid on IGRINS observations at $\Omega=6.2$ as the best fitting model in red as in Figure \ref{fig:contour}. In blue: selected model representative of the high pressure branch. In green: selected model representative of the high $G_0$ branch.  }
  \label{fig:ContourSheffer}
\end{figure}

\citet{Sheffer2011} have compared PDR models to pure rotational H$_2$ line observations with Spitzer.
Our best fit model based on IGRINS observations (cf. Section~\ref{sect:unconstrained_fits}) yields a $G_0$ value ($G_0=800$) that significantly differ from their result ($G_0=10^4$). However, we also find an L-shaped degeneracy in our parameter estimation, with high pressure and a high $G_0$ branches (as shown on Fig.~\ref{fig:ContourSheffer}). In this appendix, we thus compare the predictions of three models to the Spitzer pure rotational H$_2$ observations of \citet{Sheffer2011} (taking their observations at the emission of peak H$_2$ emission): our best fit model ($P_\mathrm{th}= 4.1\times 10^6$ K cm$^{-3}$, $G_0=800$), a model at the tip of the high $G_0$ branch ($P_\mathrm{th}= 1.5\times6$ K cm$^{-3}$, $G_0=1.8\times10^4$), and a model at the tip of the high pressure branch ($P_\mathrm{th}= 5\times10^7$ K cm$^{-3}$, $G_0=1000$) .

 \begin{figure}[htbp] 
  \centering
  \includegraphics[width=0.5\textwidth]{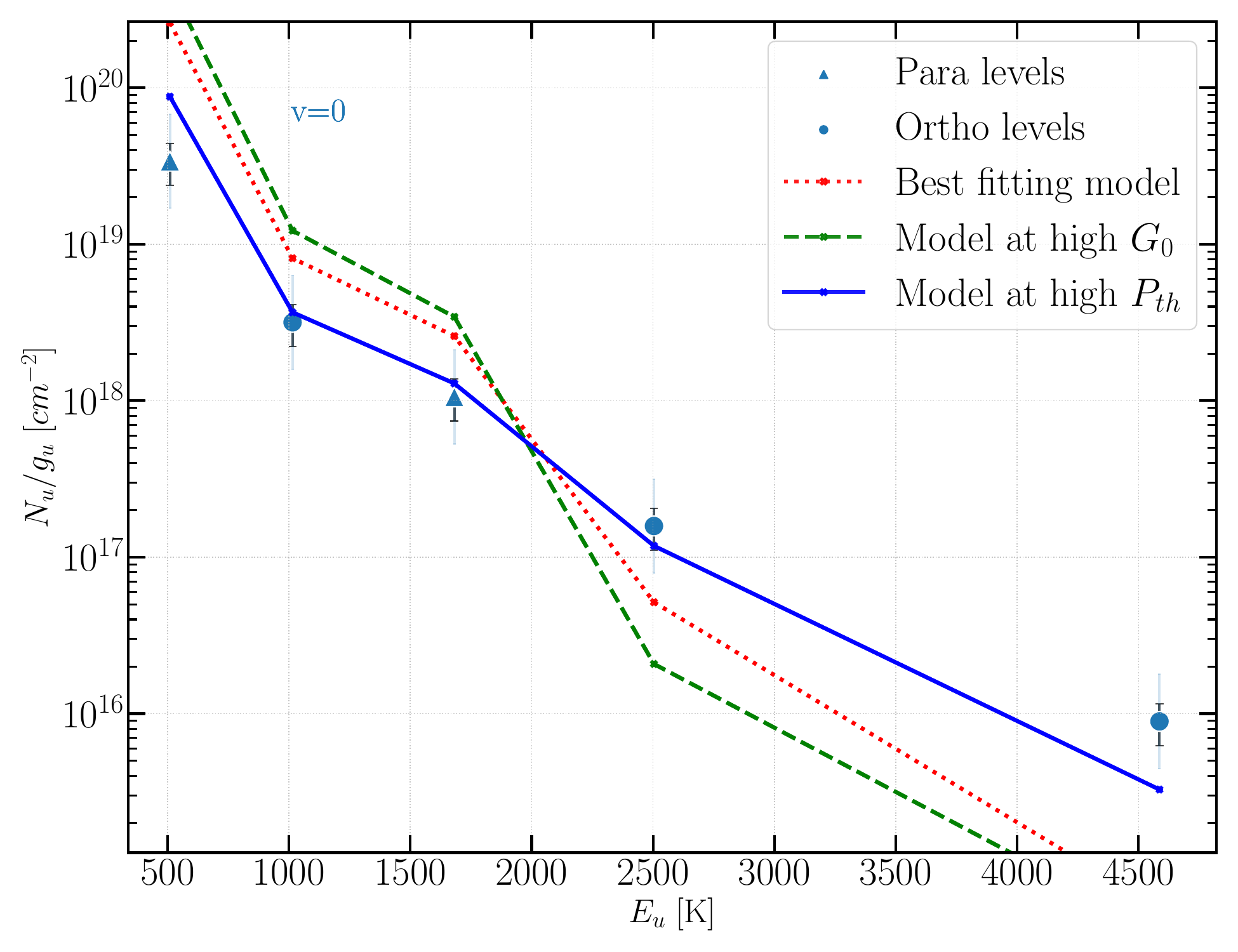}
  \caption{Excitation diagram of 3 models from the NGC 2023 grid presented in Fig.~\ref{fig:ContourSheffer} and compared to the Spitzer observation of pure-rotational H$_2$ lines at the H$_2$ emission peak position in \citet{Sheffer2011}.}
  \label{fig:Sheffer}
\end{figure}

The resulting excitation diagrams for pure rotational H$_2$ lines from \citet{Sheffer2011} compared to the three models are presented in Fig.~\ref{fig:Sheffer}.
The model within the high-pressure branch reproduces the \citet{Sheffer2011} observations significantly better than both the model that best fits the IGRINS observations in red and the model at high FUV intensity in green. 
Consequently, we conclude that the Spitzer observations favor the high-pressure branch among models compatible with the IGRINS observations.


\section{Spatial Origin and Physical Conditions of NIR H$_2$ Emission}\label{Appendix:Profile}

\begin{figure}
  \centering
  \includegraphics[width=0.48\textwidth]{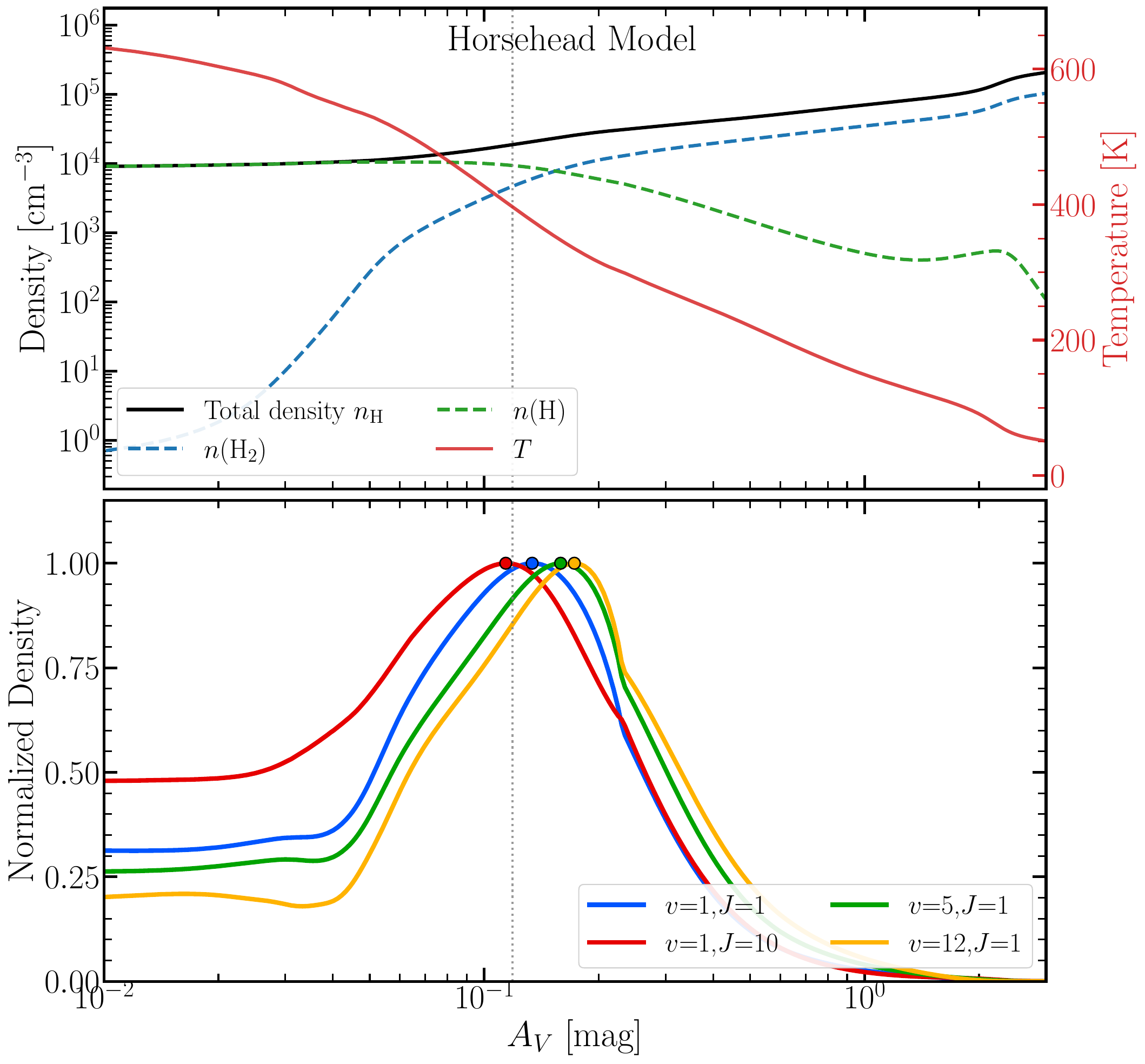}
  \includegraphics[width=0.48\textwidth]{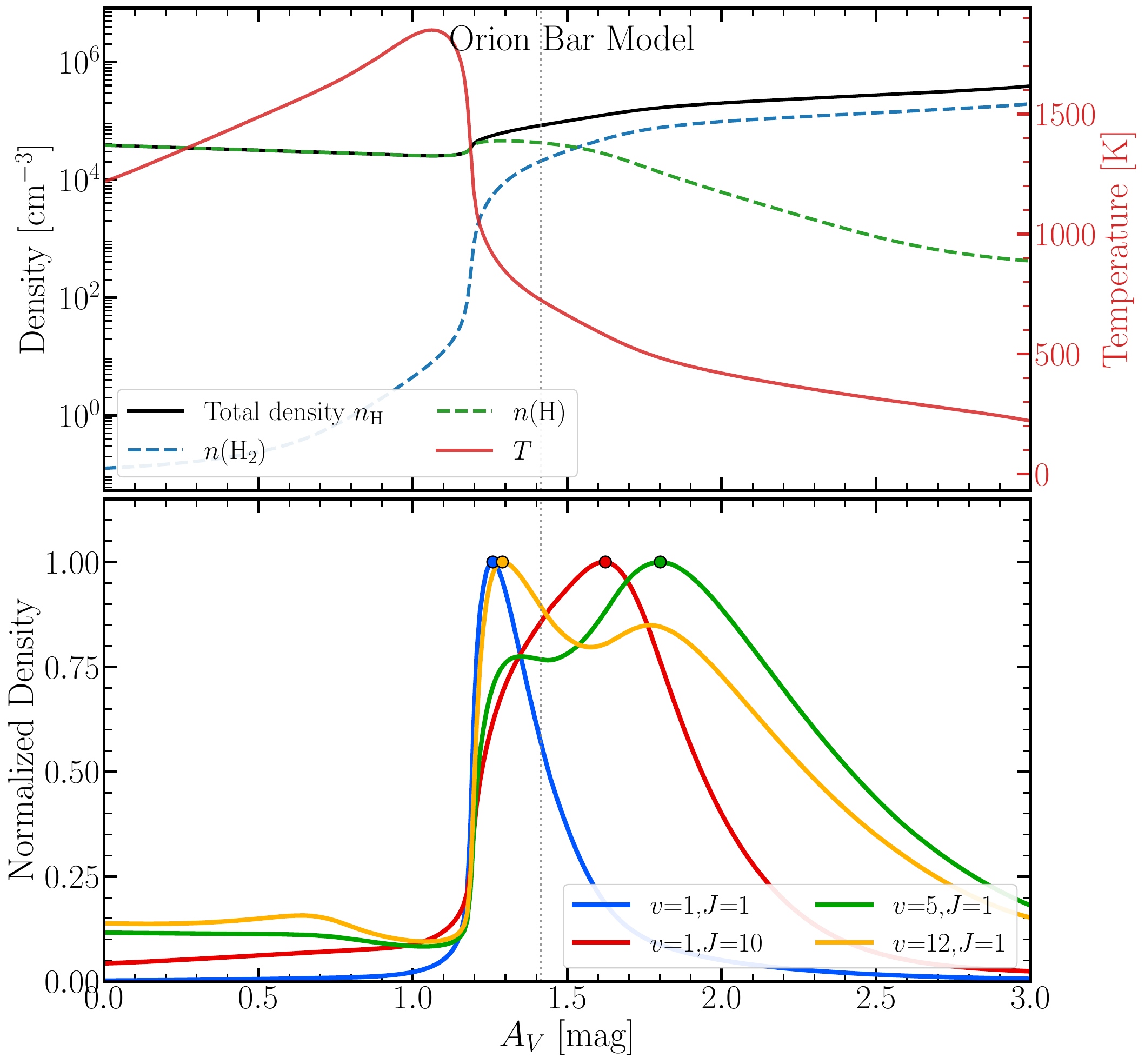}
  \caption{\rev{Physical structure and H$_2$ excitation of the A: the Horsehead Nebula model, B: the Orion Bar model. \textit{Top panel:} Profiles of total hydrogen density $n_{\mathrm{H}}$ (solid black), atomic hydrogen $n(\mathrm{H})$ (dashed green), molecular hydrogen $n(\mathrm{H}_2)$ (dashed blue), and gas temperature $T$ (solid red, right axis).  \textit{Bottom panel:} Normalized densities of selected $H_2$ rovibrational levels $(v, J)$     as a function of visual extinction $A_V$. The vertical dashed line marks the     $\mathrm{H}/\mathrm{H}_2$ transition region.}}
  \label{fig:Orion_Av}
\end{figure}

\rev{In this appendix, we present the spatial origin of the near-infrared H$_2$ lines in the Orion Bar and the Horsehead Nebula, taken as two representative PDRs. Figure~\ref{fig:Orion_Av} shows the spatial structure predicted by the best-fitting models for these two regions.}

\rev{According to the theory of the H/H$_2$ transition of \citet{Sternberg2014}, the Horsehead Nebula corresponds to a PDR with a low $\alpha G$, whereas the Orion Bar corresponds to conditions with a high $\alpha G$. In the first case, H$_2$-dissociating FUV photons are predominantly absorbed by H$_2$ Lyman and Werner transitions. The H/H$_2$ transition takes place close to the ionization front (A$_\mathrm{V} \simeq 0.1$ in our Horsehead model) and the transition is gradual, i.e. the density of H$_2$ increases smoothly with depth. To the contrary, in the case of the Orion Bar, H$_2$-dissociating FUV photons are mostly absorbed by grains so the transition takes place close to A$_\mathrm{V}$ = 1 (here 1.4). Also, contrary to the Horsehead, the transition is sharp. The gas temperature and the density at the H/H$_2$ transition, defined as $n(\mathrm{H}) = 2 \times n(\mathrm{H}_2)$, are 350 K and $7\times10^{3}$ cm$^{-3}$ in the case of the Horsehead and, 700 K and $2\times10^4$ cm$^{-3}$, in the Orion Bar.}

\rev{In the physical conditions of the Horsehead Nebula, the H$_2$ levels responsible for the ro-vibrational emission are primarily populated through UV pumping, followed by fluorescence and subsequent radiative cascades, at least for levels with excitation energies below 40,000 K (see Fig.~\ref{fig:H2_Domin_process}). The populations of these levels increase progressively as H$_2$ begins to self-shield against the incident UV radiation, and then decrease beyond the dissociation front, where most photons capable of pumping H$_2$ have been absorbed. As a result, the near-infrared lines probe a narrow region around the H/H$_2$ transition, where the gas properties vary relatively smoothly.}

\rev{In contrast, in the Orion Bar model, the populations of collisionally excited levels, such as $v=1$, $J=1$, peak ahead of the H/H$_2$ transition, where the gas remains warm. Their populations then decrease rapidly as the temperature drops across the dissociation front. Conversely, the populations of levels populated through UV pumping, fluorescence, and subsequent radiative and collisional cascades increase as soon as the H$_2$ abundance rises and remain significant up to A$_\mathrm{V} \sim 2$, i.e. as long as UV photons capable of pumping H$_2$ are present. As a result, the populations of some of these levels, such as $v=5$, $J=1$ and $v=12$, $J=1$, exhibit a double-peaked profile in the Orion Bar. This behaviour arises from the competition between two opposing effects: the UV pumping rate decreases with depth into the cloud, while the gas density simultaneously increases. Consequently, the near-infrared lines probe a more extended region around the H/H$_2$ transition than in moderate PDRs. This region is characterized by steep gradients in both temperature and density. For example, the gas temperature decreases from about 1500 K to a few hundred kelvin across the region contributing to the observed emission.}

\section{Ro-vibrational distribution of H$_2$ at formation}\label{App:Sizun}

\begin{figure}[ht]
    \centering
      \includegraphics[width=0.5\textwidth]{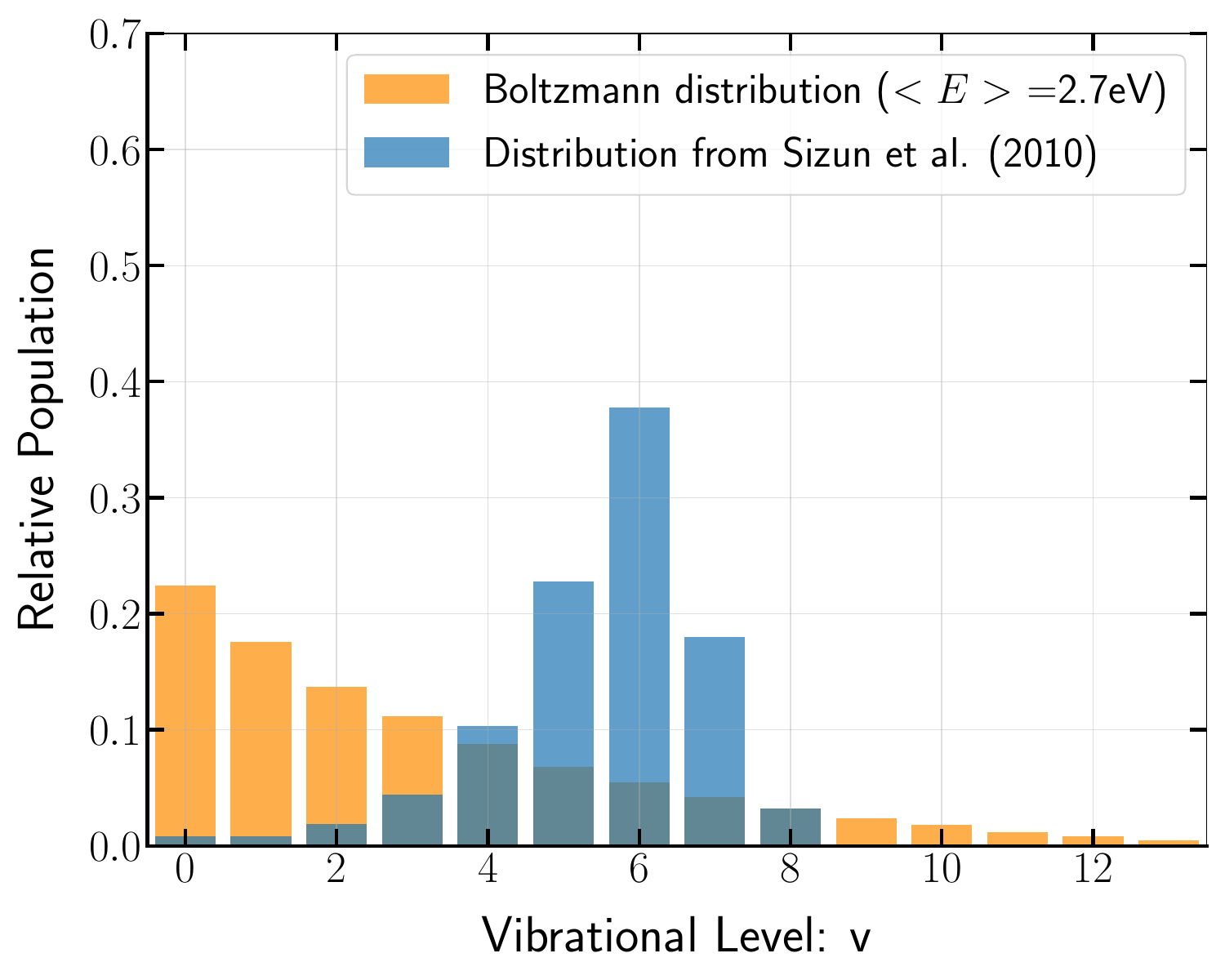}
      \includegraphics[width=0.5\textwidth]{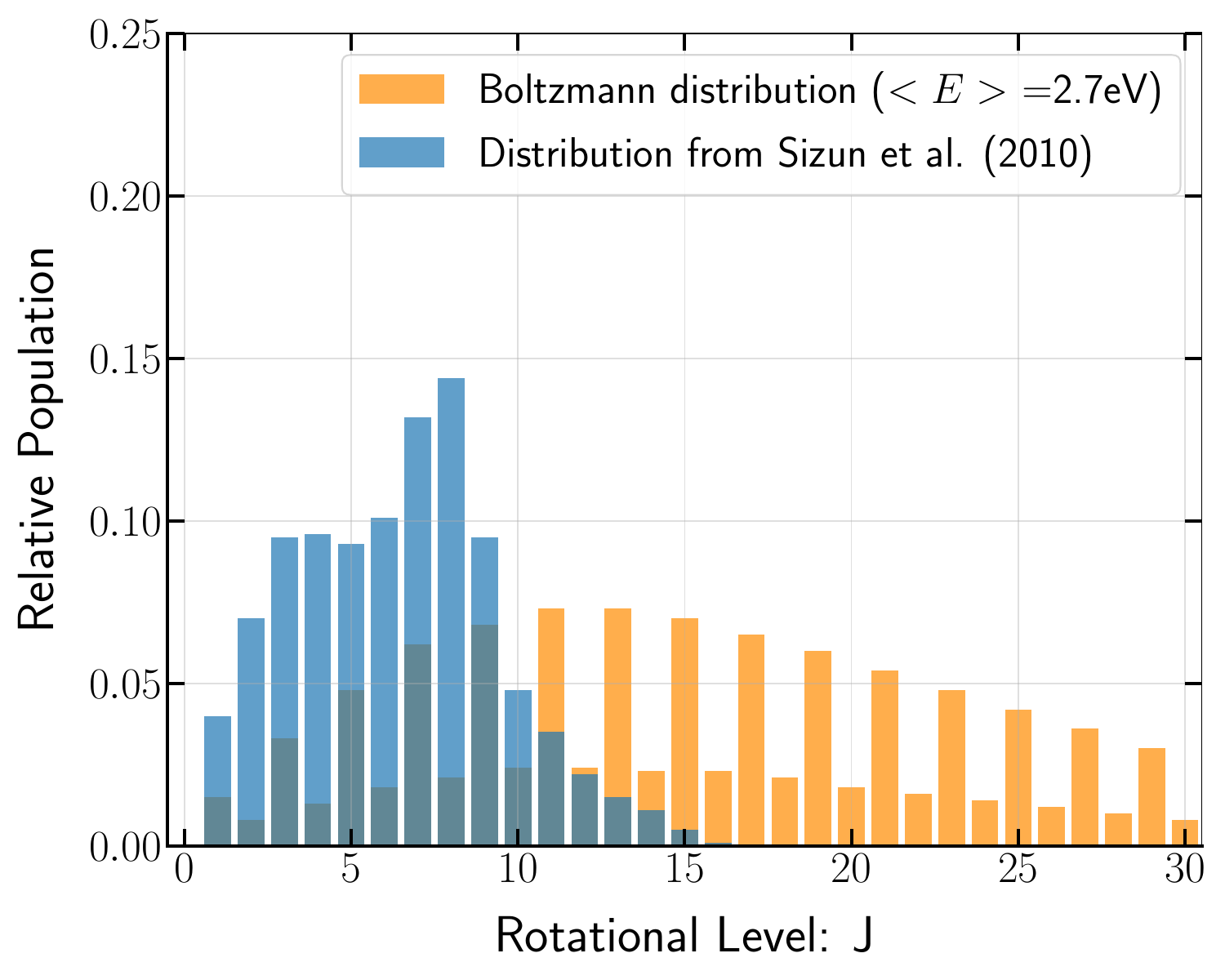}
      \caption{H$_2$ distribution at formation in $v$ and $J$ levels. In orange: the Boltzmann distribution with its mean energy corresponding to 2.7eV, in blue, the distribution presented in \citet{Sizun2010}}
  \label{fig:SizunDistrib}
 \end{figure}

In Section~\ref{Section:Sizun}, we discussed possible scenarios for the formation of H$_2$ in its rovibrational levels. We investigated the presence of signatures of H$_2$ excitation at formation under two scenarios: a Boltzmann distribution and the distribution proposed by \citet{Sizun2010} presented in Figure \ref{fig:SizunDistrib}. 

\begin{figure}[ht]
    \centering
      \includegraphics[width=0.5\textwidth]{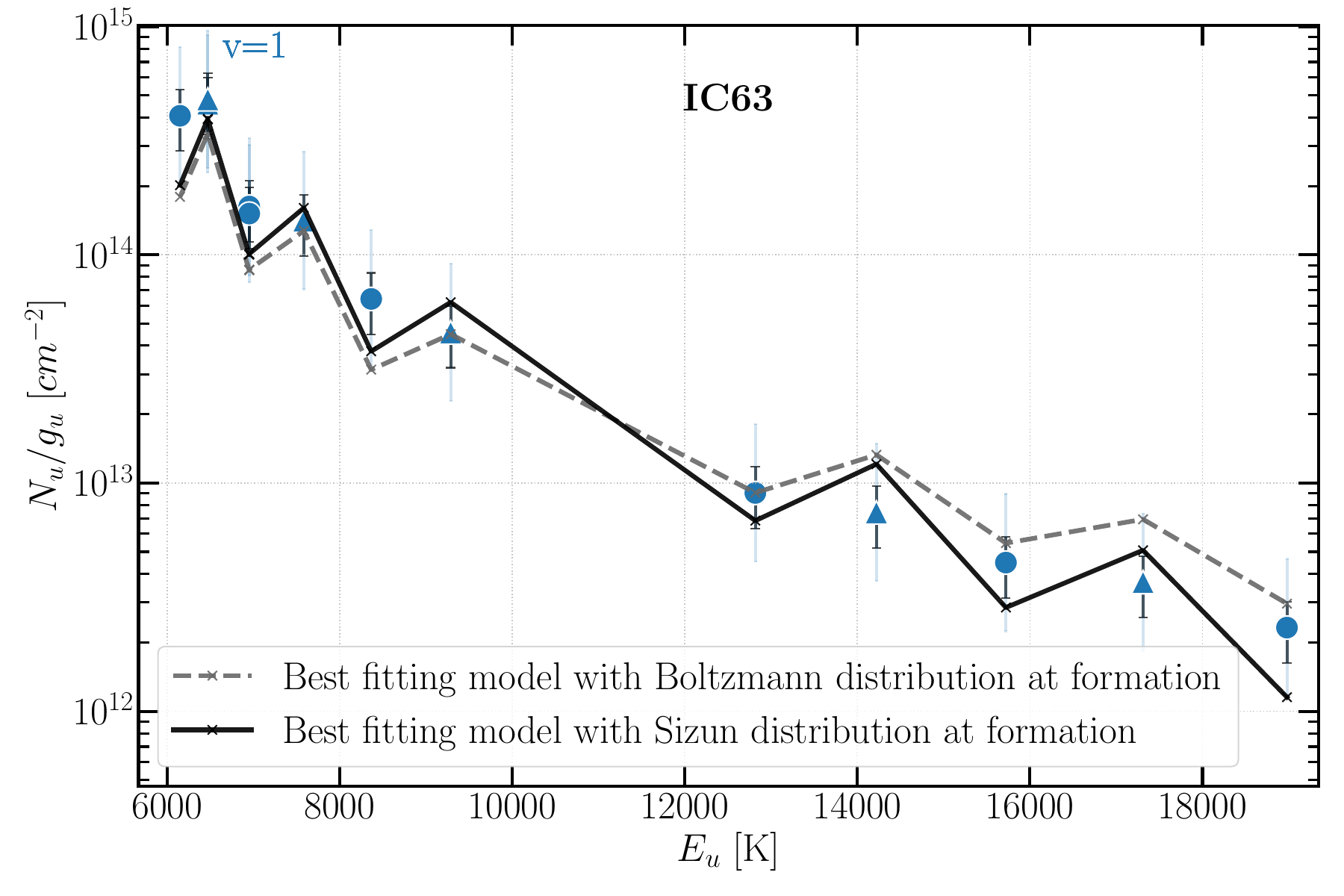}
            \includegraphics[width=0.5\textwidth]{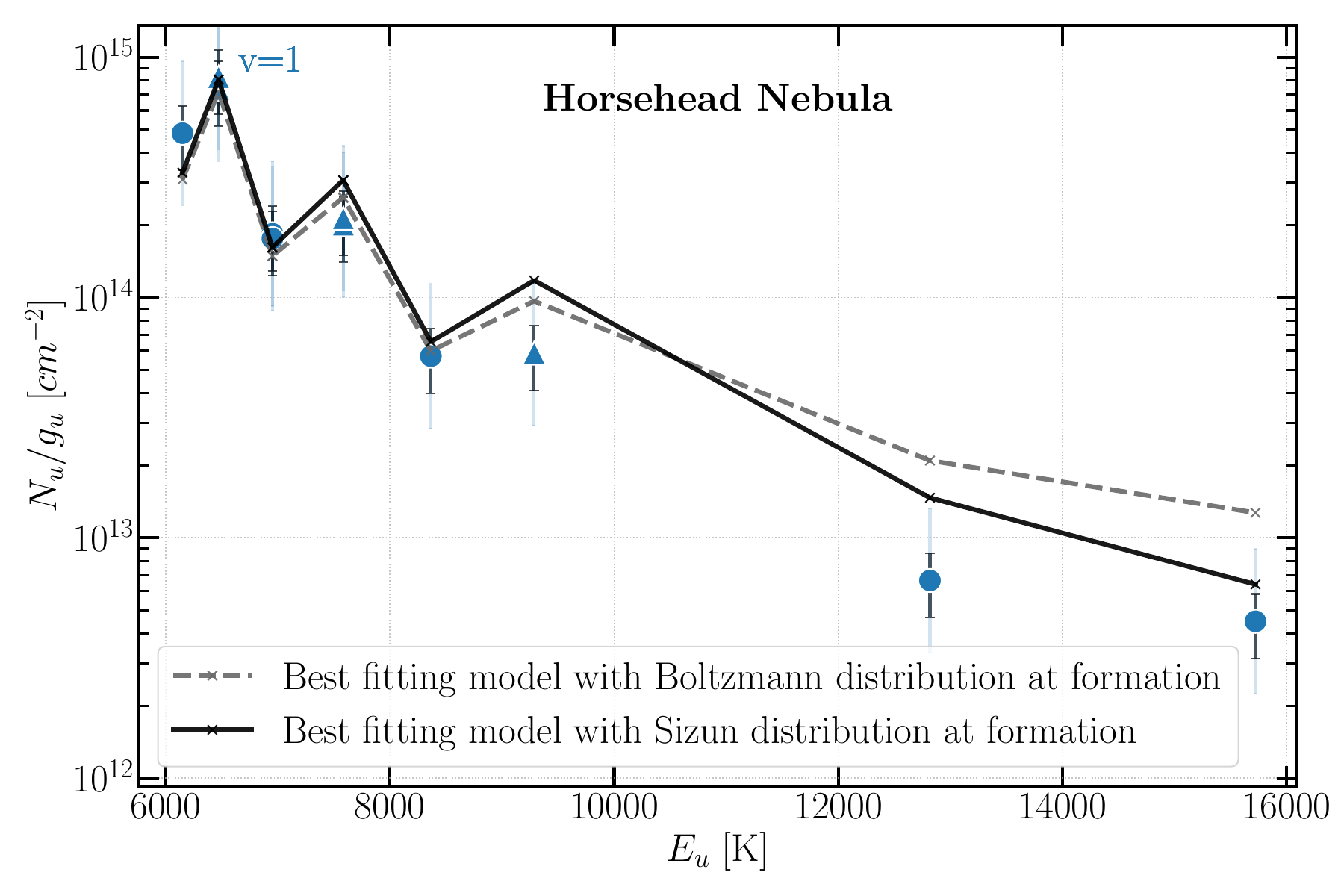}
                  \includegraphics[width=0.5\textwidth]{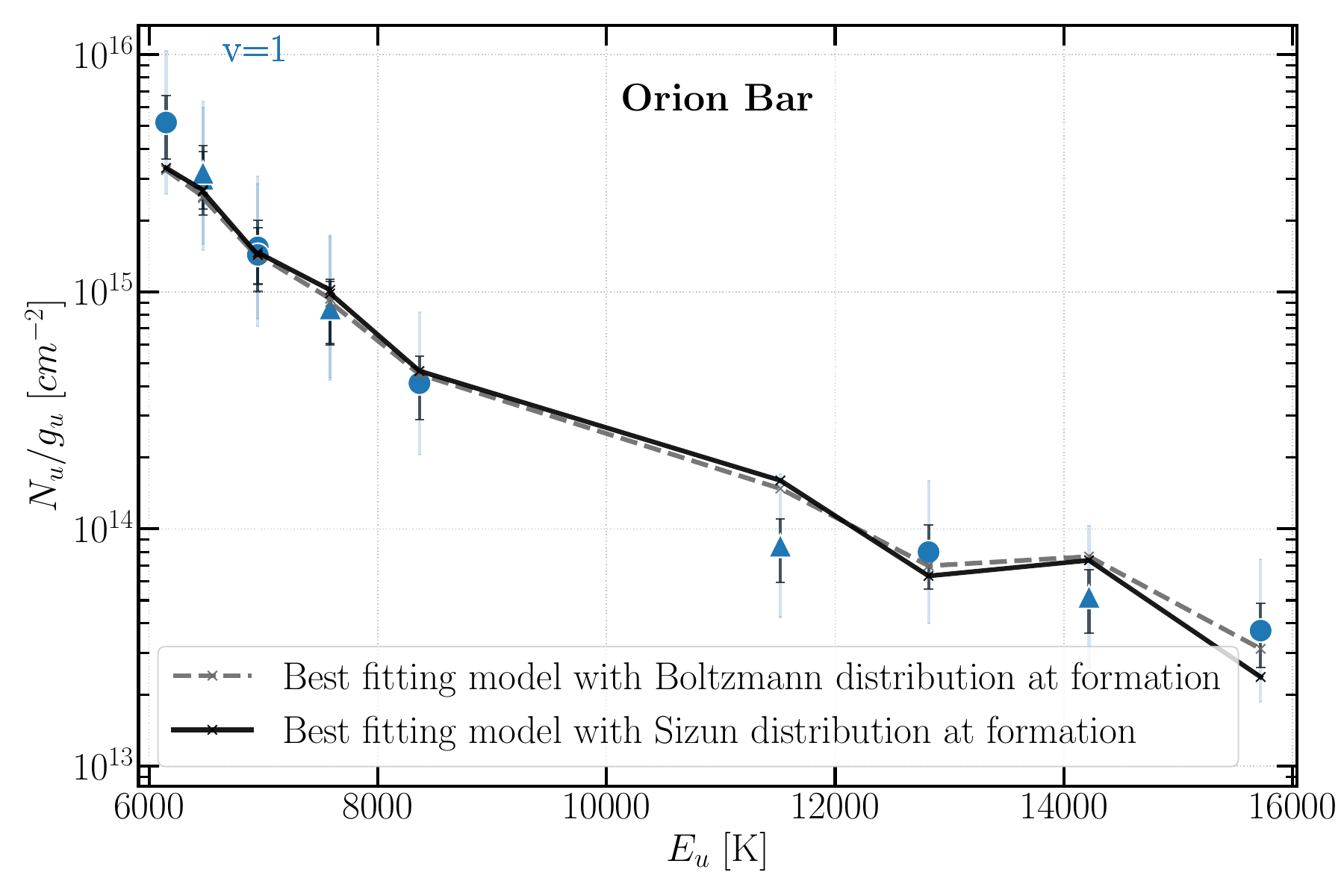}
      \caption{Same as Figure~\ref{fig:Sizun} for : IC63, NGC2023 and the Orion Bar.}
  \label{fig:SizunOthers}
 \end{figure}

In the main text, results are presented for the Horsehead and NGC2023 in Figure \ref{fig:Sizun}. Figure \ref{fig:SizunOthers} shows the comparison between the models using these two prescriptions for the other PDRs studied in this paper: IC~63, the Horsehead Nebula, and the Orion Bar.

\section{OPR by triplets for intermediate PDRs}\label{OPR}

\begin{figure*}[h]
    \centering
      	\includegraphics[width=0.9\textwidth]{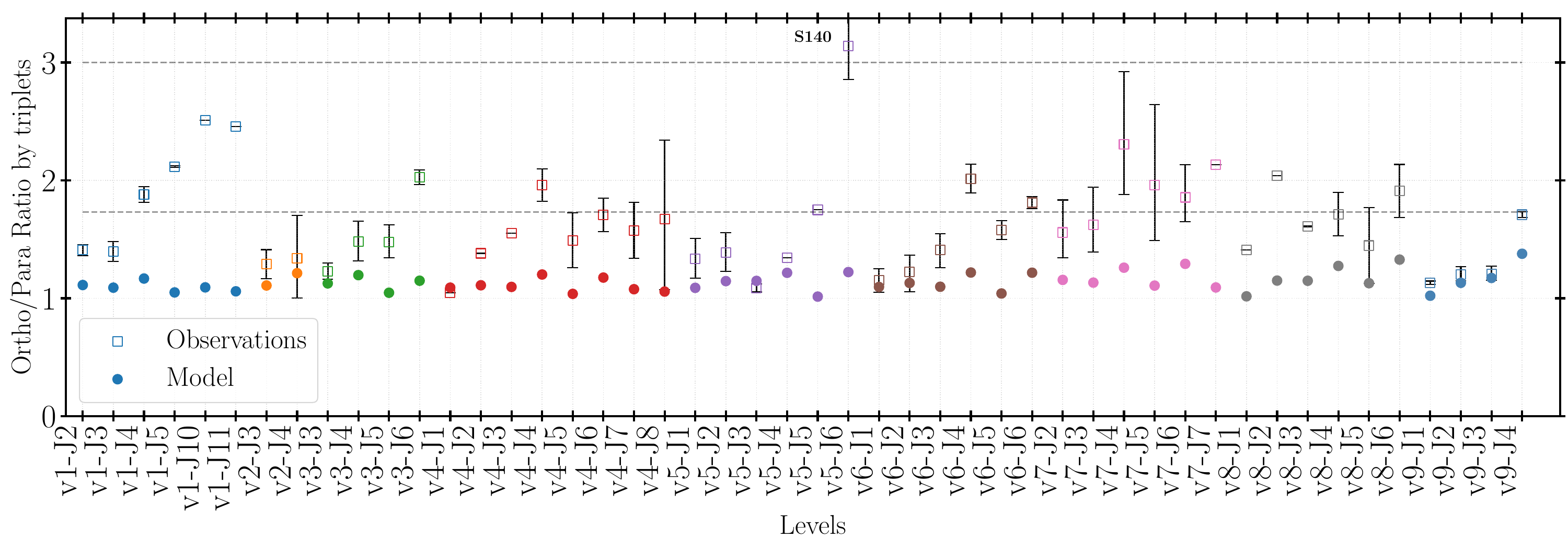}
      	\includegraphics[width=0.9\textwidth]{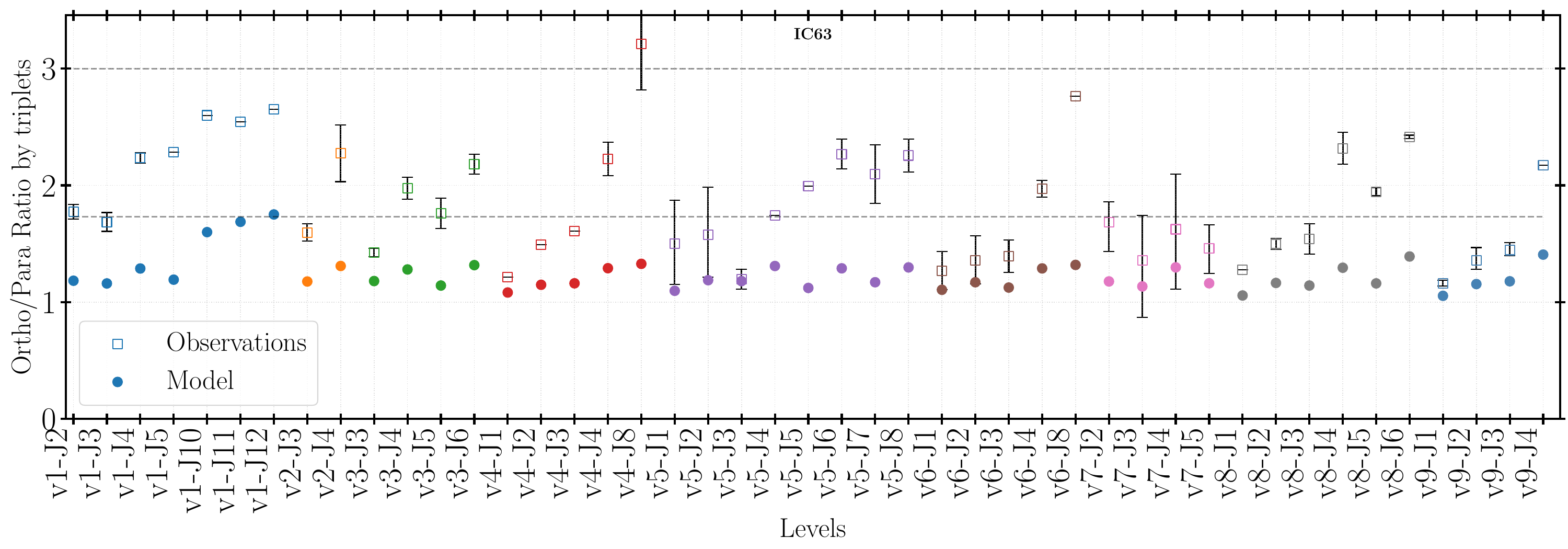}
      	\includegraphics[width=0.9\textwidth]{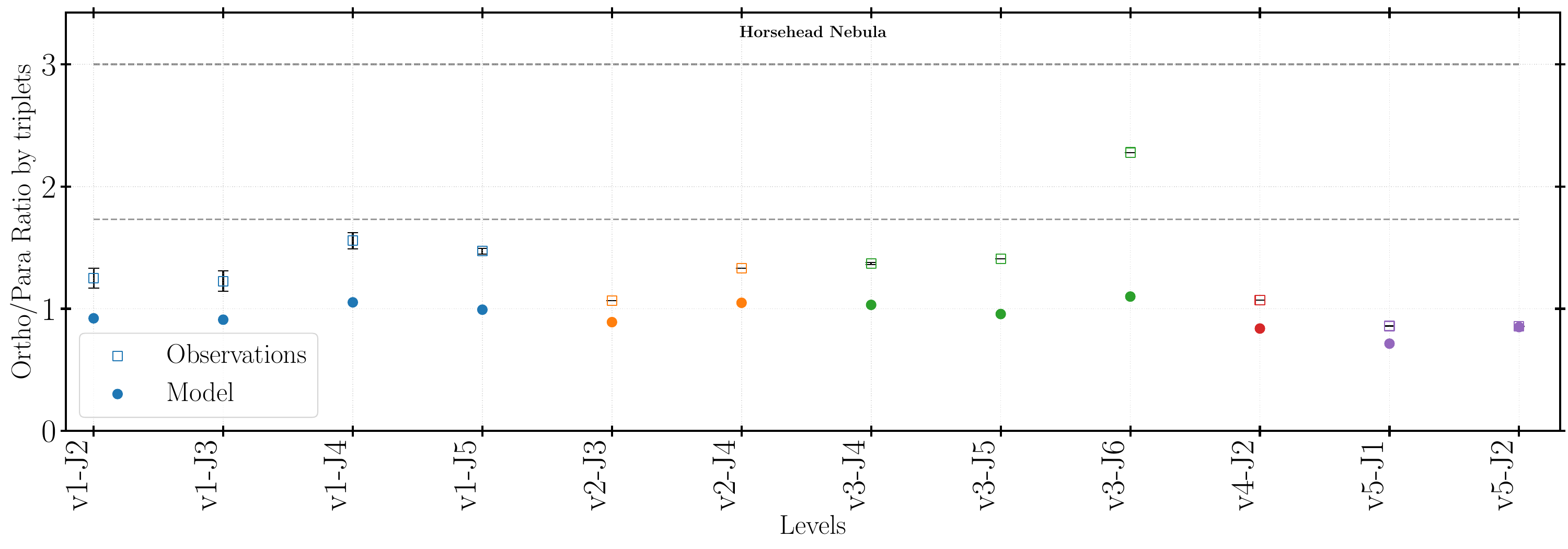}
      	\includegraphics[width=0.9\textwidth]{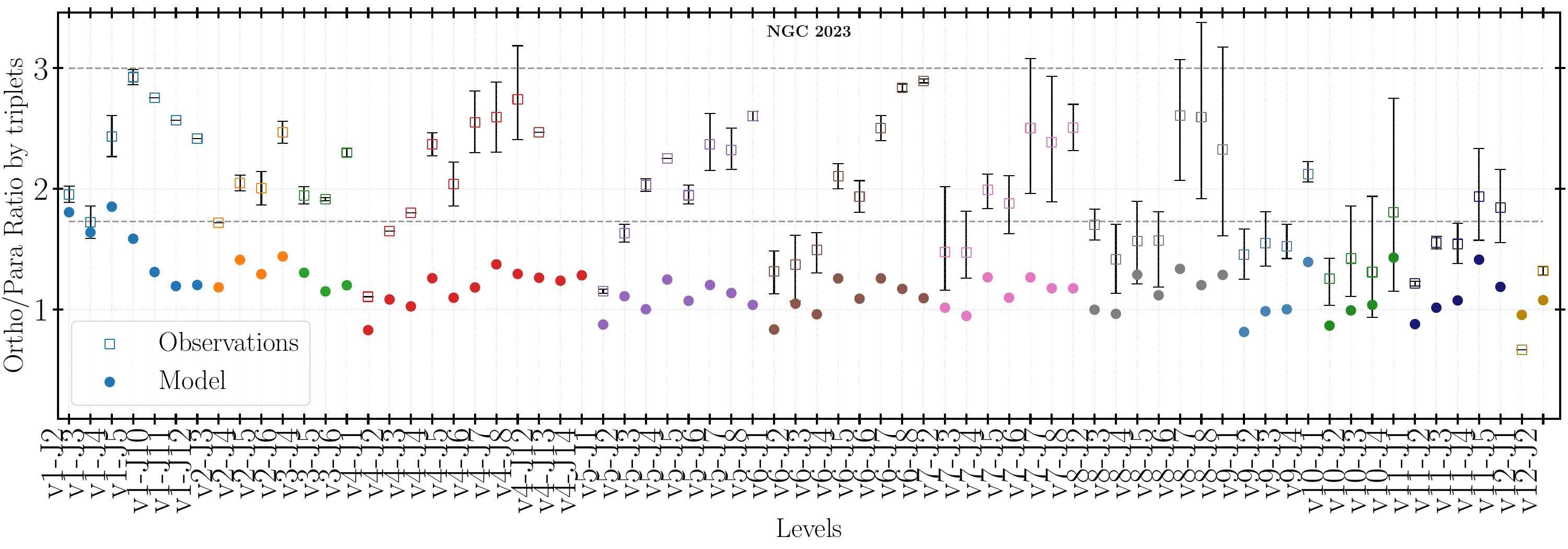}

      \caption{OPR by triplets for (a): S140, (b): IC63, (c): Horsehead Nebula, (d): NGC2023 as developed in~\ref{Section:OPR}.}
  \label{fig:OPRIntermediate}
 \end{figure*}

In Section~\ref{Section:OPR}, we discussed the ortho-para ratio estimated for each level triplets  following the methodology described in \citet{Bron2016} for every PDRs. In the main text, only the results for the Orion Bar PDR are displayed. We show here in Fig \ref{fig:OPRIntermediate} the three low excitation PDRs and NGC2023, the intermediate case between the two regimes. 

The modeled OPRs are close to the observed ones for the Horsehead PDR. For the other two low excitation PDRs, a relatively good agreement is found for the low Js of each v, but observations seem to show an OPR that then increases for higher Js, a trend that is not reproduced by the models.

In the case of NGC2023, the intermediate case between low excitation and high excitation PDRs, 
the behavior is more similar to that highlighted in the Orion Bar case (see main text), with a model OPR slightly above 1, and observed OPRs that are close to the model results for low Js but quickly rise towards 3 for higher Js.

\clearpage


\end{document}